\documentclass[11pt,a4paper]{article}
\pdfoutput=1
\usepackage{jinstpub}

\usepackage{xspace} 
\usepackage[usenames,dvipsnames]{color}
\usepackage{booktabs,graphicx,mathrsfs,verbatim,amsmath,units,soul}
\usepackage{ulem}
\usepackage{xspace} 
\usepackage{amsmath} 
\usepackage{upgreek}
\usepackage{gensymb}
\usepackage{tabularx}
\usepackage{fullwidth}
\usepackage{multirow}
\usepackage [autostyle, english = american]{csquotes}
\usepackage{subcaption}
\usepackage{graphicx}
\usepackage{xcolor}
\MakeOuterQuote{"}

\newcolumntype{Y}{>{\centering\arraybackslash}X}

\newcommand{\RNum}[1]{\uppercase\expandafter{\romannumeral #1\relax}}

\begin{document}

\title{Proposal of a Geiger-geometry Single-Phase Liquid Xenon Time Projection Chamber as Potential Detector Technique for Dark Matter Direct Search}

%\linenumbers

%\author{M.~Anthony}\columbia
%\author{E.~Aprile}\columbia
%\author{L.~Grandi}\chicago
%\author{Q.~Lin}\columbia
%\author{R.~Saldanha}\pnnl

\author[a, b, 1]{Qing~Lin%
\note{Corresponding author.}}
% \author[c]{K. Ni}

\affiliation[a]{State Key Laboratory of Particle Detection and Electronics, University of Science and Technology of China, Hefei 230026, China}
\affiliation[b]{Department of Modern Physics, University of Science and Technology of China, Hefei 230036, China}
% \affiliation[c]{Department of Physics, University of California San Diego, La Jolla, CA 92093, USA}


\emailAdd{qinglin@ustc.edu.cn}

\abstract{
Dual phase time projection chamber using liquid xenon as target material is one of most successful detectors for dark matter direct search, 
and has improved the sensitivities of searching for weakly interacting massive particles by almost five orders of magnitudes in past several decades.
However, it still remains a great challenge for dual phase liquid xenon time projection chamber to be used as the detector in next-generation dark matter search experiments ($\sim$ 50 tonne sensitive mass), in terms of reaching sufficiently high field strength for drifting electrons, and sufficiently low background rate.
Here we propose a single phase liquid xenon time projection chamber with detector geometry similar to a Geiger counter, as a potential detector technique for future dark matter search, which trades off field uniformity for less isolated charge signals.
Preliminary field simulation and signal reconstruction study have shown that such single phase time projection chamber is technically feasible and can have sufficiently good signal reconstruction performance for dark matter direct search.
}

%Use showkeys class option if keyword
\maketitle

\section{Introduction}
\label{sec:introduction}
% state-of-the-art for dark matter direct search
Dual phase liquid xenon (LXe) Time Projection Chamber (TPC) has been gaining popularity in the field of dark matter direct search during past several decades, thanks to its capability of simultaneously reconstructing scintillation and ionization signals of one event and event vertex.
Experiments using such detector technique, like PandaX-I/-II~\cite{xiao2015low, tan2016dark}, XENON100/1T~\cite{aprile2012dark, baudis2018dark},  and LUX~\cite{akerib2017results}, have been able to keep refreshing the world record of sensitivities for high-mass ($>$6\,GeV) weakly interacting massive particle (WIMP) search.
Currently, second-generation dark matter direct search experiments using dual phase LXe-TPC with multi-tonne sensitive mass, such as PandaX-4T~\cite{zhang2019dark}, XENONnT~\cite{aprile2020projected}, and LZ~\cite{akerib2020projected}, are under commissioning and soon will start to collect scientific data, exploring parameter space that is unreachable by previous detectors.
% challenges for dual phase TPC to be 3rd-generation detector
However, it remains unclear if dual phase TPC can reach sufficiently good signal quality and ultra-low background level that are required in third-generation dark matter search experiment ($\sim$50\,tonne sensitive mass)~\cite{aalbers2016darwin}, to search for dark matter particles that have weaker interaction rate with baryonic matter than current experimental precision.
Especially in the sense of reaching high cathode voltage to provide enough drift field strength and of mitigating high isolated ionization background rate, R\&D efforts are needed.

% single-phase status and drawback of dual phase
On the contrary to the extensive attentions that dual-phase TPCs have received, single phase TPCs have not raised much interest in the field of dark matter direct search.
Although single-phase TPC using liquid argon as material has been widely deployed nowadays in large-scale neutrino experiments, such as MicroBooNE~\cite{acciarri2017design}, ICARUS~\cite{acciarri2015proposal}, and DUNE~\cite{abi2018dune}, there is no existing dark matter search experiment utilizing such technique.
Unlike dual phase TPC which utilize electroluminescence in gas to convert charge signals into proportional light signals and collect them, traditional single phase TPC collects charge signals through wire planes~\cite{acciarri2015proposal}, thus has a higher charge collection threshold that is not sufficiently low for sensitive dark matter direct search.
% recent measurements of proportional scintillation in liquid xenon
Electroluminescence in LXe opens up new possibilities for single phase LXe TPC that can reach low charge collection threshold.
It was first demonstrated by~\cite{lansiart1976development} and confirmed by~\cite{masuda1979liquid} a few years later.
There are measurements~\cite{arazi2013first, aprile2014measurements, ye2014initial, juyal2020proportional} recently, showing that the proportional light signals in LXe start to become visible when the field strength reaches 412\,kV/cm.
The estimated electron amplification is comparable to the one obtained through electroluminescence in gaseous xenon in a traditional dual phase LXe TPC.
It is worth noting that TPC with micro-pattern anode (such as thick-GEM~\cite{breskin2013liquid} and MicroMegas~\cite{giomataris1996micromegas}) and uniform electric drift field is also an option for single phase TPC.
However the disadvantage of such option is that micro-pattern anode requires relatively large amount of materials to be placed in detector, which can reduce the light collection of TPC and increase background rate.

% layout of the paper
In this manuscript, we propose a single phase LXe TPC which collects charge signals through electroluminescence in LXe and has different geometry than traditional dual phase TPC currently widely used in dark matter direct search experiments. 
Such TPC with new geometry trades off field uniformity for low background rate caused by isolated charge signals.
And it is proposed mainly as possible detector for next-generation dark matter direct search, but also may have potential broad application in neutrino measurement or medical imaging.
In Section~\ref{sec:detector}, a review of dual phase TPC and a description of proposed Geiger-geometry single phase TPC are given.
In Section~\ref{sec:field_simulation} and~\ref{sec:signal_simulation}, we show the results of field simulation and study of signal reconstruction for such proposed single-phase TPC, with an assumed prototype detector diameter of 3\,meter.
%Two prototypes with different detector sizes are studied in this work, one small-size prototype with detector diameter of 20\,cm and one large-size prototype with diameter of 3\,m.
% The large-size prototype is comparable to next-generation dark matter search detector.
% The small-size prototype is presented as reference for future R\&D work.

\section{TPCs for Next-Generation Dark Matter Search}
\label{sec:detector}

Dual phase TPC is currently used in second-generation dark matter search experiments, such as PandaX-4T~\cite{zhang2019dark}, XENONnT~\cite{aprile2020projected}, and LZ~\cite{akerib2020projected}, which are under commissioning.
It is also the default detector technique that is going to be used in future 3rd-generation dark matter search experiments, like DarkSide-20k~\cite{aalseth2018darkside}, DARWIN~\cite{aalbers2016darwin}, and PandaX-30T~\cite{liu2017current}.
Although dual phase TPC has shown excellent signal responses in past experiments, next-generation dual phase TPCs will face non-trivial challenges since they will reach unprecedented size (sensitive mass of about several ten tonne) in the field of dark matter direct search, especially it may face challenges in mitigating isolated ionization signals causing accidental pileup.
The major backgrounds of a TPC-based dark matter search detector consist of four components: electronic recoil (ER) background, neutron background, surface background, and accidental coincidence (AC) background (see~\cite{collaboration2019xenon1t} for more details).
As the detector gets larger, ER background which is due to diffused radioactive impurities is reduced since concentration of impurities is diluted.
Also various techniques were developed in recent years to increase the removal rate of radioactive impurities (such as cryogenic distillation~\cite{cui2020design, aprile2017removing, baudis2017online}).
Most neutrons scatter multiple times in large detector, and can be rejected as the detector has position reconstruction capability.
Both neutron and surface backgrounds can be significantly rejected by fiducialization.
But the isolated charge signal rate grows as detector size gets bigger, increasing the rate of AC background which is uniformly distributed in fiducial volume.
%Various R\&D studies are ongoing to demonstrate the feasibility of dual phase TPC in next-generation experiments.
On the other hand, single phase TPC which collects charge through electroluminescence in LXe is another possible technique that can be used in next-generation detector.
The electroluminescence in LXe has been observed and the amplification factor is comparable with the one obtained through electroluminescence in gaseous xenon.

\subsection{Dual Phase Time Projection Chamber}

\begin{figure}[htp]
\centering
\includegraphics[width=0.95\textwidth]{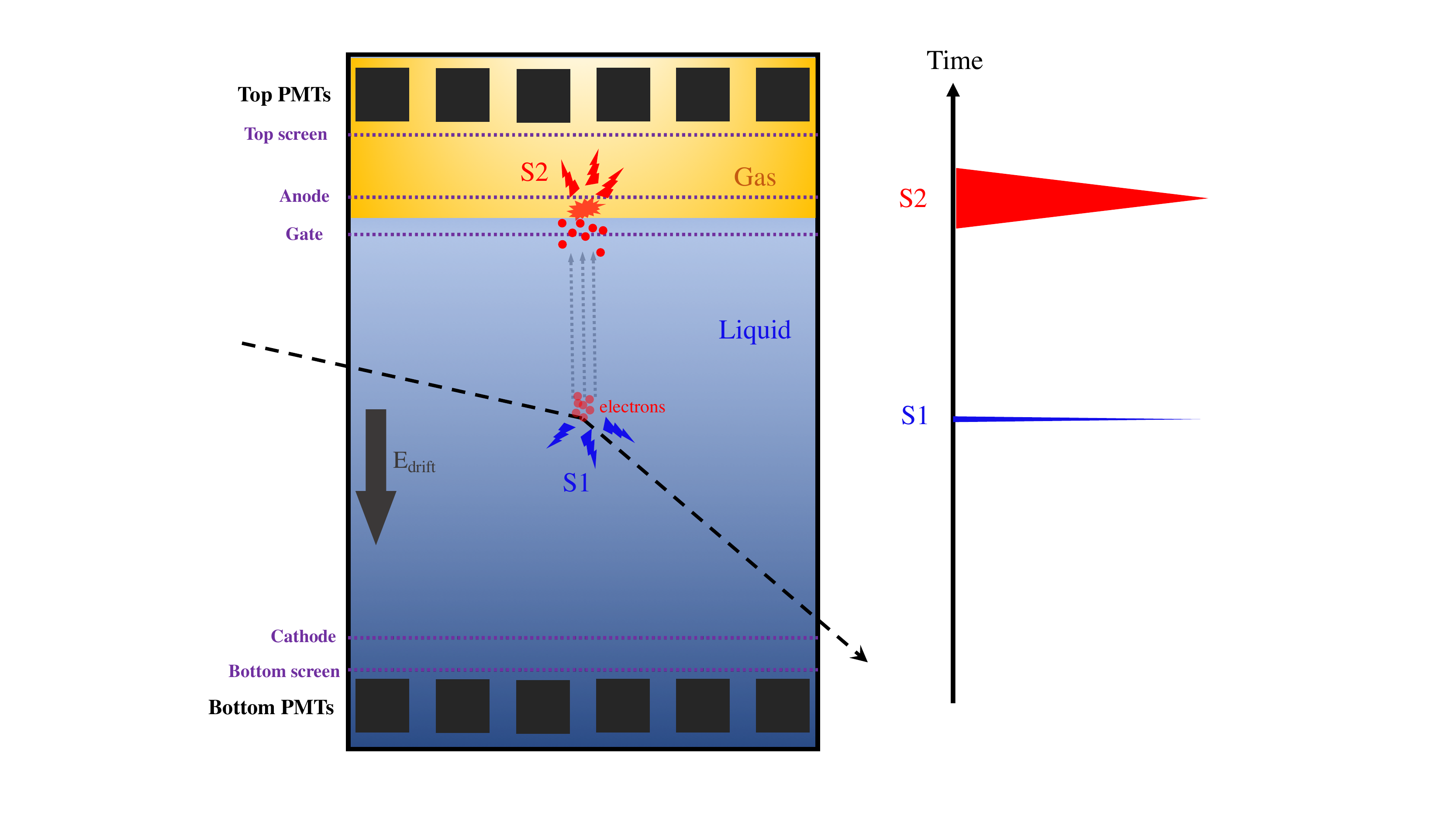}
\caption{Diagram of a typical dual phase TPC.}
\label{fig:dual_phase_tpc}
\end{figure}

% detector description 
Dual phase TPC consists of mainly three parallel plane electrodes, light reflectors and light sensors.
PTFEs are traditionally used as light reflectors, and photo-multipliers (PMTs) as light sensors placed at top and bottom of cylindrical TPC sensitive volume. 
Fig.~\ref{fig:dual_phase_tpc} shows a typical dual phase TPC diagram.
Three parallel plane electrodes are (from bottom to top) cathode, gate, and anode.
Usually gate electrode is grounded, negative and positive voltages are applied to cathode and anode, respectively.
Cathode voltage is typically in the order of several ten kV to provide a drift field strength of several hundred to thousand kV/cm in the sensitive volume.
Distance between anode and gate is typically $\sim$5\,mm, and liquid surface is placed between them.
Anode voltage is required to be sufficiently strong so that the field strength in gas is above the threshold for electroluminescence ($\sim$1.3\,kV/cm/atm~\cite{aprile2004proportional} in gaseous xenon and $\sim$0.55\,kV/cm/atm~\cite{monteiro2011determination} in gaseous argon) and gives enough efficiency for extracting electrons out of liquid surface (100\% extraction efficiency at $\sim$10\,kV/cm in gas for xenon~\cite{aprile2014observation}).
Because of the high voltages applied to cathode and anode and their proximity to the top and bottom PMTs, additional electrode planes (named screen electrodes) are needed in front of top and bottom PMTs for their protection.

% Signal mechanism & Pros of dual phase TPC
The sensitive volume in a dual phase TPC is a cylinder between cathode and gate planes. 
Deposit energy in sensitive volume is converted into prompt scintillation light signals and ionization electrons.
Prompt light signals are detected by PMTs, and called S1 conventionally.
Ionization electrons are drifted towards the proportional amplification region which is between gate and anode planes. 
Under stronger field in amplification region, electrons are extracted into gas and produce proportional light signals which are detected by PMTs and named S2.
S1 signal has fast time response in the scale of several ten to hundred nanoseconds, determined by the singlet and triplet decay constant of excimer of target element.
On contrary, S2 signal has relatively slow time response in scale of several microseconds, which is dominated by the electron diffusion during drifting and the electron traveling in gas gap after liquid surface extraction.
Therefore, S1 and S2 signals are distinguishable on recorded event waveform.
The ability to reconstruct the scintillation and ionization signals of an energy deposition is the key of dual phase TPC's success in dark matter direct search.
Amplitudes of S1 and S2 signals, as well as the pulse shape information of S1 signal in certain material like argon, can provide discrimination power between ERs and nuclear recoils (NRs), and can further reduce the effective backgrounds induced by gamma and betas. 
The ER rejection power can reach about 10$^{-10}$~\cite{gorel2014search} in liquid argon, using pulse shape discrimination.
Several large-size experiments ($>$100\,kg active mass) using LXe as target material have reached an ER-NR discrimination power (defined as leakage fraction with 50\% NR acceptance) ranging from about 8$\times$10$^{-4}$ to 5$\times$10$^{-3}$~\cite{akerib2020discrimination, collaboration2018signal, collaboration2018dark, xiao2014first, cui2017dark}, but depending on field strength, detector geometry, and energy.
In addition, time information of S1 and S2 signals and S2 signal's pattern on PMTs can be used to reconstruct the longitude and transverse positions, respectively, giving 3-D position reconstruction of interaction vertex.
The longitude position reconstruction resolution can typically reach about 3\,mm, dictated by the width of S2 signals.
The transverse resolution can range from 0.8\,cm to 1.5\,cm~\cite{aprile2019xenon1t, akerib2018position} in the region of interest (ROI) for dark matter search.
This also provides extra rejection power against background from external or material radiations through fiducialization, and against neutron backgrounds through discrimination between single and multiple scatters.

% Cons of dual phase TPC and explicit challenges that next-generation experiments are facing
The excellent reconstructions of S1, S2 and event vertex requires high level of field uniformity in both the drift and amplification regions.
On the other hand, high transparency of electrode plane is critical in order to maintain a sufficiently high light collection efficiency to search for dark matter signals.
This demands the electrode plane to be made of fine wires and maintain good tension on them under cryogenic temperature, which by itself is a great challenge to multi-ten-tonne dual phase TPC.
On top of this, it is not feasible to apply very high voltage to anode in a dual phase TPC since the anode is located in gas and close to top PMTs (easy to discharge).
To maintain adequate field strength in dual phase TPC, high negative voltage is applied on cathode.
Because of their proximities with bottom PMTs, cathode wires may have surface field large enough so that the probability of electrons from electrode metal being extracted out through Fowler-Nordheim effect~\cite{bodnia2021electric} is non-trivial.
Extracted electrons can produce proportional scintillation through electroluminescence in liquid material under very strong field near cathode wire, increasing the background rate and limiting the voltage applied to cathode.
% \textcolor{red}{(Think of a way to find the clue of this in past experiments.)}
Besides, existence of large amount of metal electrodes and liquid surface in dual phase TPC is very likely one of the major causes for large amount of isolated ionization signals (events without a detected over-threshold scintillation signal) in past experiments (such as in XENON100~\cite{aprile2014observation}). 
Photoionization on metal electrodes produces single- or multi-electron emission, and liquid surface may trap part of the drifted electrons and release them with a time delay~\cite{sorensen2017electron, sorensen2017two}. 
The issue of isolated ionization signals can become even more critical in next-generation detector as the detector scales up.

\subsection{Geiger-geometry Single Phase Time Projection Chamber}
\begin{figure}[htp]
\centering
\includegraphics[width=0.95\textwidth]{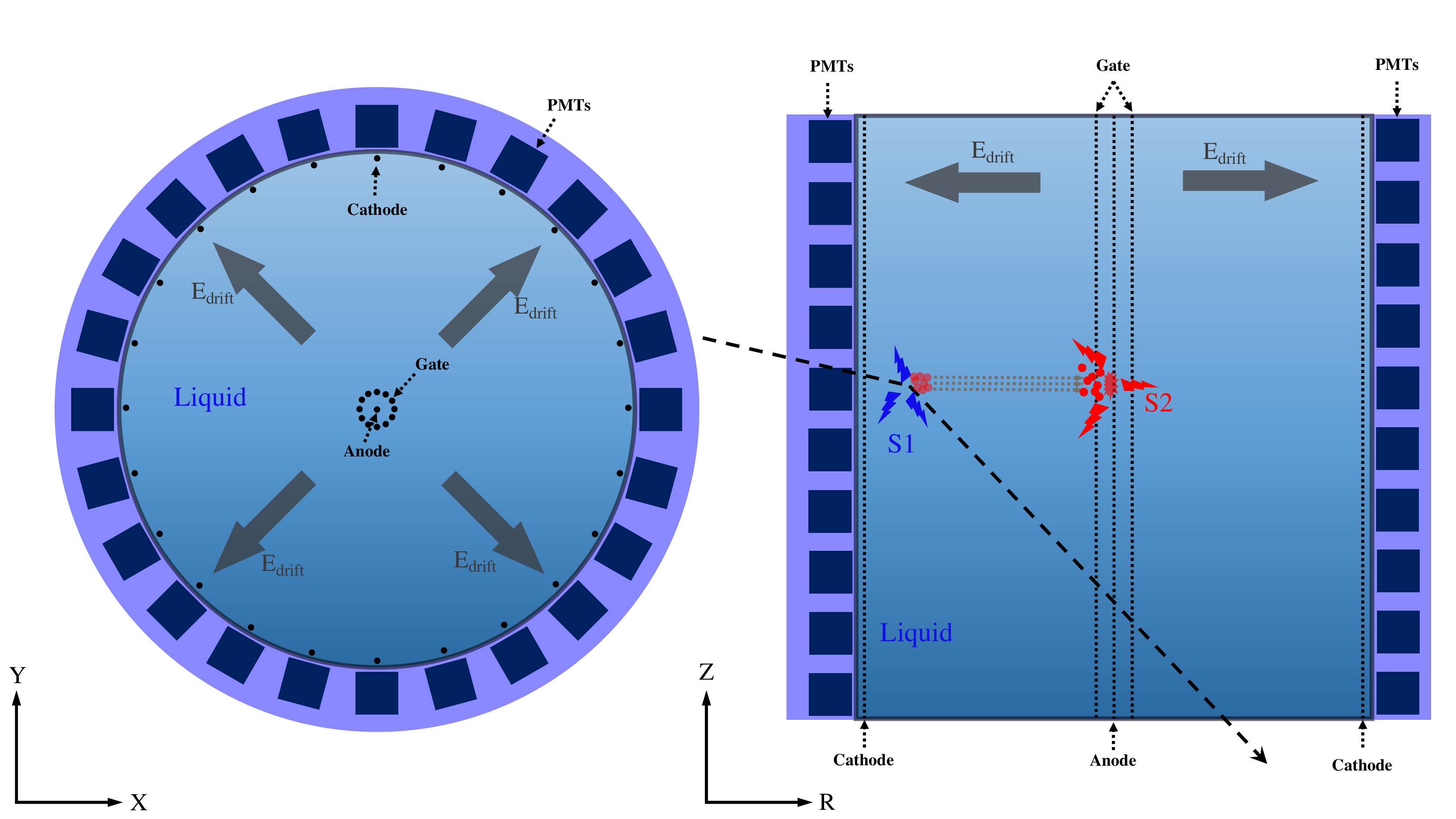}
\caption{
Conceptual diagram of the proposed Geiger-type single phase TPC.
Left is the top view, and right plot is the side view.
}
\label{fig:single_phase_tpc}
\end{figure}

% single phase TPC description
We propose a new single phase TPC with detector geometry similar to a Geiger counter. 
It collects scintillation light signals through instrumented light sensors, same as dual-phase TPC.
But it amplifies charge signals through electroluminescence in LXe, instead of in gaseous xenon in a dual phase TPC.
The simplest conceptual diagram of the proposed Geiger-geometry single phase TPC can be seen in Fig.~\ref{fig:single_phase_tpc}.
Geiger-Geometry single phase Time Projection Chamber (GG-TPC) has also a cylindrical sensitive volume. 
Unlike the parallel electrode planes in a dual phase TPC, GG-TPC has parallel wires serving as electrodes.
A single wire at the central axis of GG-TPC's cylinder serves as anode, surrounded by several wires with fixed spacing as gate wires to regulate the amplification fields.
Wires at the cylindrical side of GG-TPC's sensitive volume, parallel to the central axis and equally spaced between adjacent wires, serve as cathode.
Light sensors, such as PMTs, can be instrumented to cover the surface of GG-TPC's cylindrical side.

% Single mechanism and Pros and cons of single phase
The signal detection mechanism of GG-TPC is similar to that of a dual phase TPC, but the ionized electrons are drifted to the central axis of sensitive volume, and go through proportional amplification near anode wire.
Such GG-TPC geometry can achieve a field larger than what is needed for electroluminescence in LXe near anode wire of reasonable macroscopic size (unlike the fine 10\,$\mu$m- and 5\,$\mu$m-wires used in~\cite{aprile2014measurements} with a traditional detector geometry), and in the meantime can also maintain moderate field strengths in the active volume between cathode and gate.
The most important information is the radial and zenith positions of an event which is required to perform fiducialization of GG-TPC.
The radial position in GG-TPC can be reconstructed using the drift time, and the zenith position may be reconstructed using the hit pattern of S2 on light sensors.
The angular position is not critical since detector geometry is axial symmetrical.
%The S2 hit pattern may have ``dark zones'' because the proportional lights are produced very near anode wire, aiding the reconstruction of angular position to some extent.
Compared to the dual phase TPC, GG-TPC can have anode and gate on positive high voltage because both of them are immersed in liquid.
Thus, the cathode voltage can be much smaller and still maintain moderate field strength in most part of sensitive volume, avoiding ultra-high field strength between cathode and PMTs.
The region close to cathode may have low drift field strength and even field distortion, as well as high background rate, and will be removed in analysis.
Also, mechanically it is much easier to maintain tension on wires in GG-TPC than wires on electrode planes in dual phase TPC.
Absence of liquid surface and less electrode material used can bring less isolated charge signal in GG-TPC, reducing AC backgrounds.
However, field strength in GG-TPC along radial axis is changing and non-uniform which may bring challenges for reconstructions of energy and radial position. 
More studies of signal reconstruction in GG-TPC are in Section~\ref{sec:signal_simulation}.
% Dedicate calibrations using injected sources, such as $^{83m}$Kr~\cite{kastens2009calibration}, with uniform event distribution expected are required to correct the effect brought by non-uniform drift field.

\section{Drift and Amplification Fields of GG-TPC}
\label{sec:field_simulation}

\begin{figure}[htp]
\centering
\includegraphics[width=0.95\textwidth]{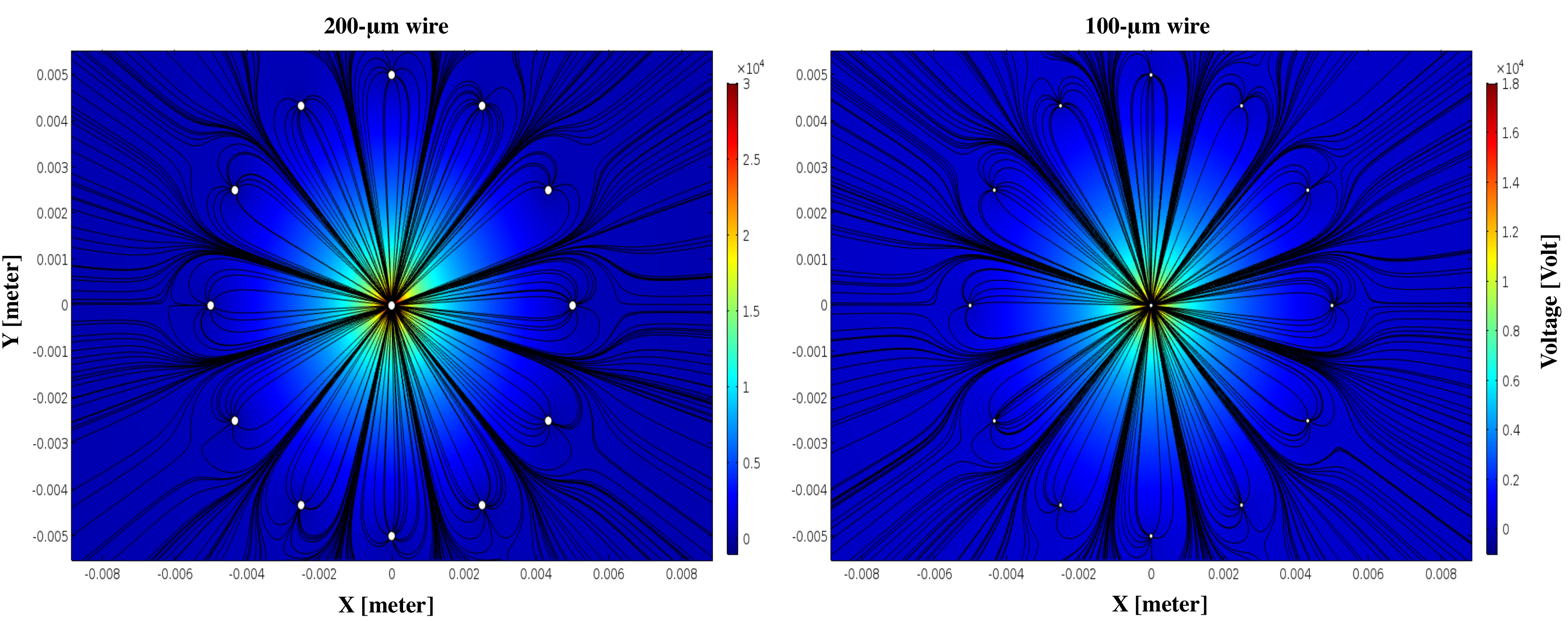}
\caption{
Simulated field lines and voltages in gate-anode region for the prototype GG-TPC. 
Left plot is for wire with diameter of 200\,$\mu$m and right plot for wire with diameter of 100\,$\mu$m.
The color axes show the voltage at different position.
}
\label{fig:sim_field_line}
\end{figure}

% simulation setup 
We have performed field simulations using COMSOL~\cite{multiphysics1998introduction}, to estimate the field configuration of GG-TPC.
% \sout{The first prototype GG-TPC (GG-TPC-P1) is assumed to have a cylindrical sensitive volume with radius of 10\,cm and height of 30\,cm, with 36 cathode wires and 12 PMTs instrumented.}
The prototype GG-TPC is assumed to have cylindrical sensitive volume with radius of 3\,meter and height of 3\,meter.
450 cathode wires and 100 PMTs are instrumented.
Twelve gate wires, equally spaced between adjacent wires, are placed 5\,mm away from the central anode wire.
Two simulations of local electric field between gate and anode are performed with different wire diameters, 100\,$\mu$m and 200\,$\mu$m, both of which are commercially available wire sizes and have been extensively used in past dark matter search experiments.
Wires made by stainless steel or tungsten with such diameters are anticipated to be strong enough, to tolerate the tension caused by cryogenic shrinking of large-scale GG-TPC.
In most commonly used operation mode, cage surface of PMT is set at negative voltage (e.g., about -800\,V for R8520 1'' Hamamatsu PMT~\cite{aprile2012measurement}).
In order to have minimal amount of electrode wires, there is no screen electrode wires to be placed in front of PMTs in GG-TPC.
Therefore, the cathode voltage can not be too high for the protection of the PMTs.
In simulations of prototype GG-TPC, the cathode voltage is fixed at -1\,kV and the gate at +10\,kV.

\begin{figure}[htp]
\centering
\includegraphics[width=0.48\textwidth]{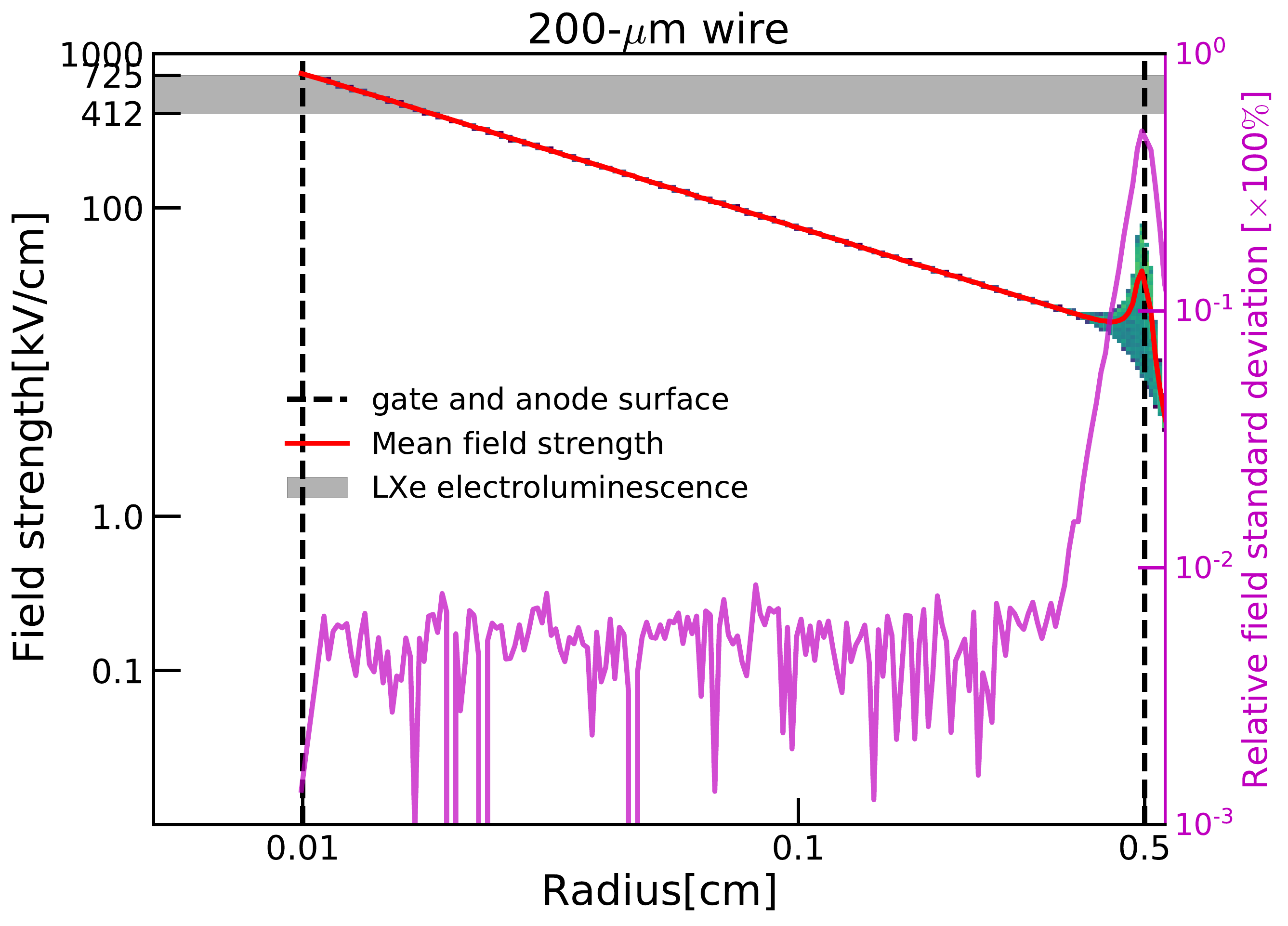}
\includegraphics[width=0.48\textwidth]{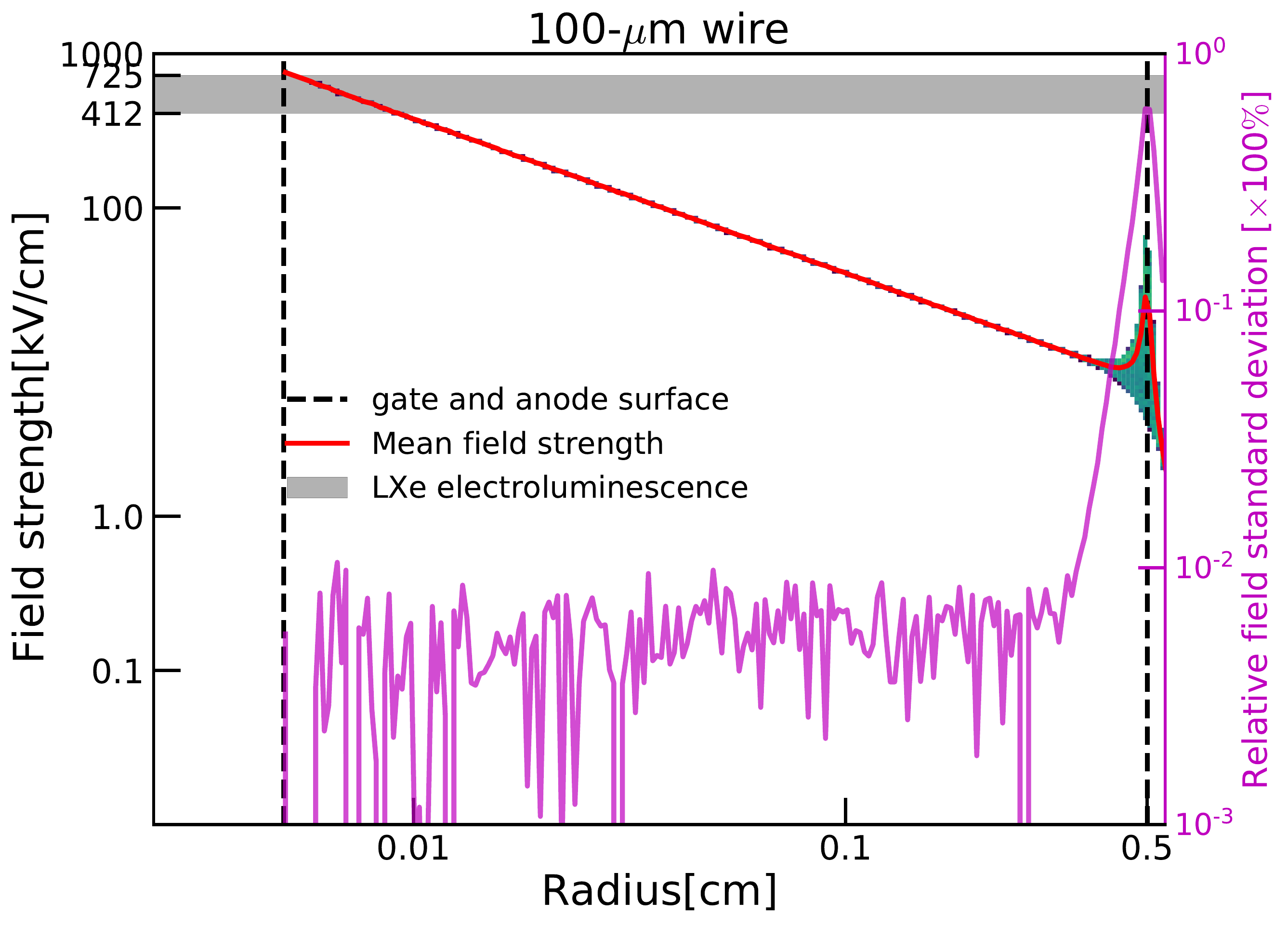}
\caption{
Distribution of GG-TPC volume on field strength versus radial position in gate-anode region. 
The red and magenta solid lines give the mean and standard deviation of field strength, respectively, at different radial position.
The vertical dashed lines indicate the position of anode surface and gate wires, respectively.
% The radius of GG-TPC-P1 is 10\,cm.
The left and right plots are for prototype GG-TPC with 200-$\mu$m and 100-$\mu$m wires, respectively.
The shaded regions indicate the allowed field range for electroluminescence in LXe~\cite{aprile2014measurements}.
% In the results, gate wires are grounded and cathode wires are set to -1\,kV.
% +30\,kV and +18\,kV are applied to anode wires for GG-TPC-P1 with 200-$\mu$m and 100-$\mu$m wires, respectively.
Differences of voltages applied to gate and anode are 30 and 18\,kV for prototype GG-TPCs with 200-$\mu$m and 100-$\mu$m wires, respectively.
}
\label{fig:sim_field_vs_R}
\end{figure}

\begin{figure}[htp]
\centering
\includegraphics[width=0.48\textwidth]{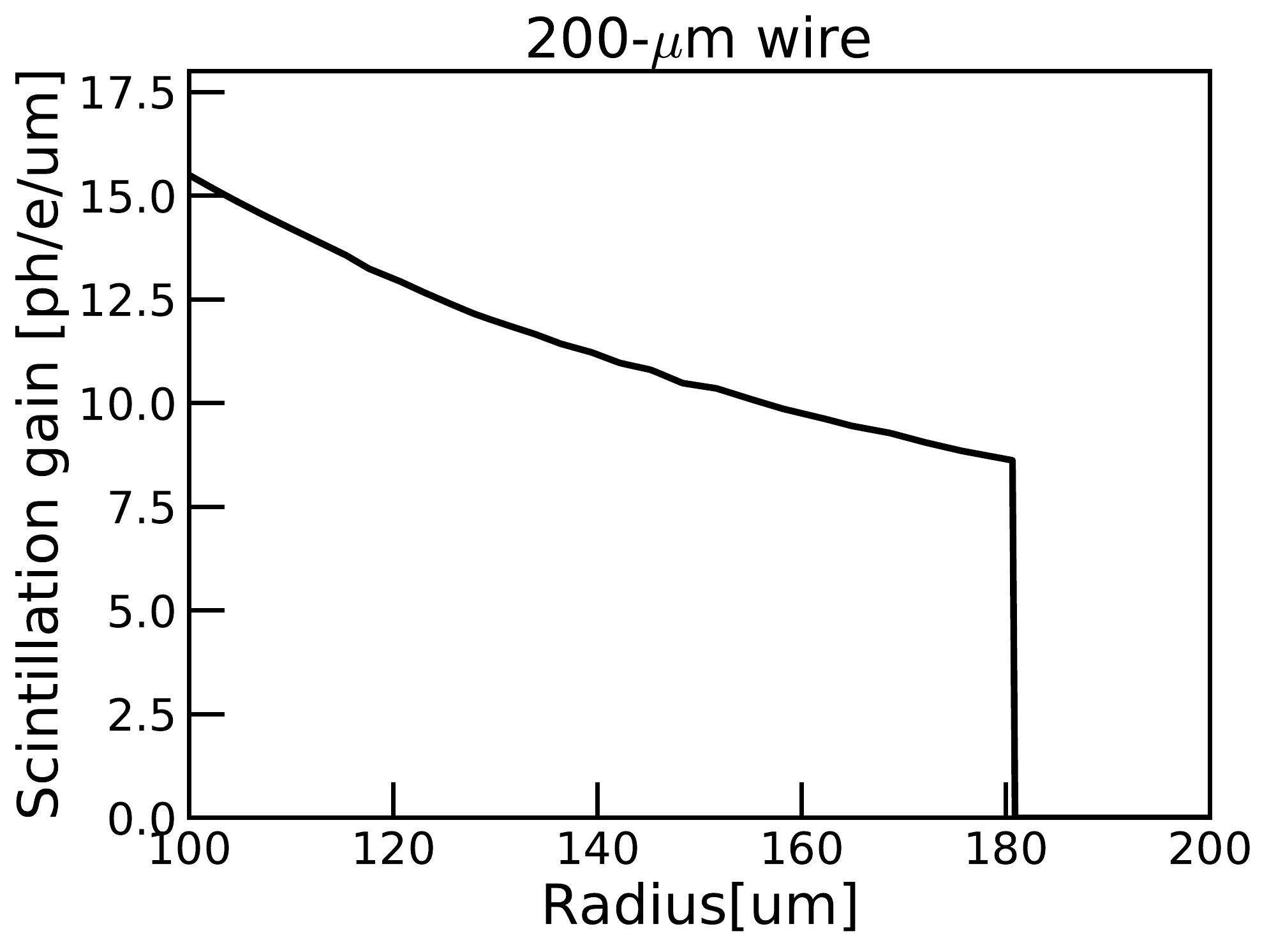}
\includegraphics[width=0.48\textwidth]{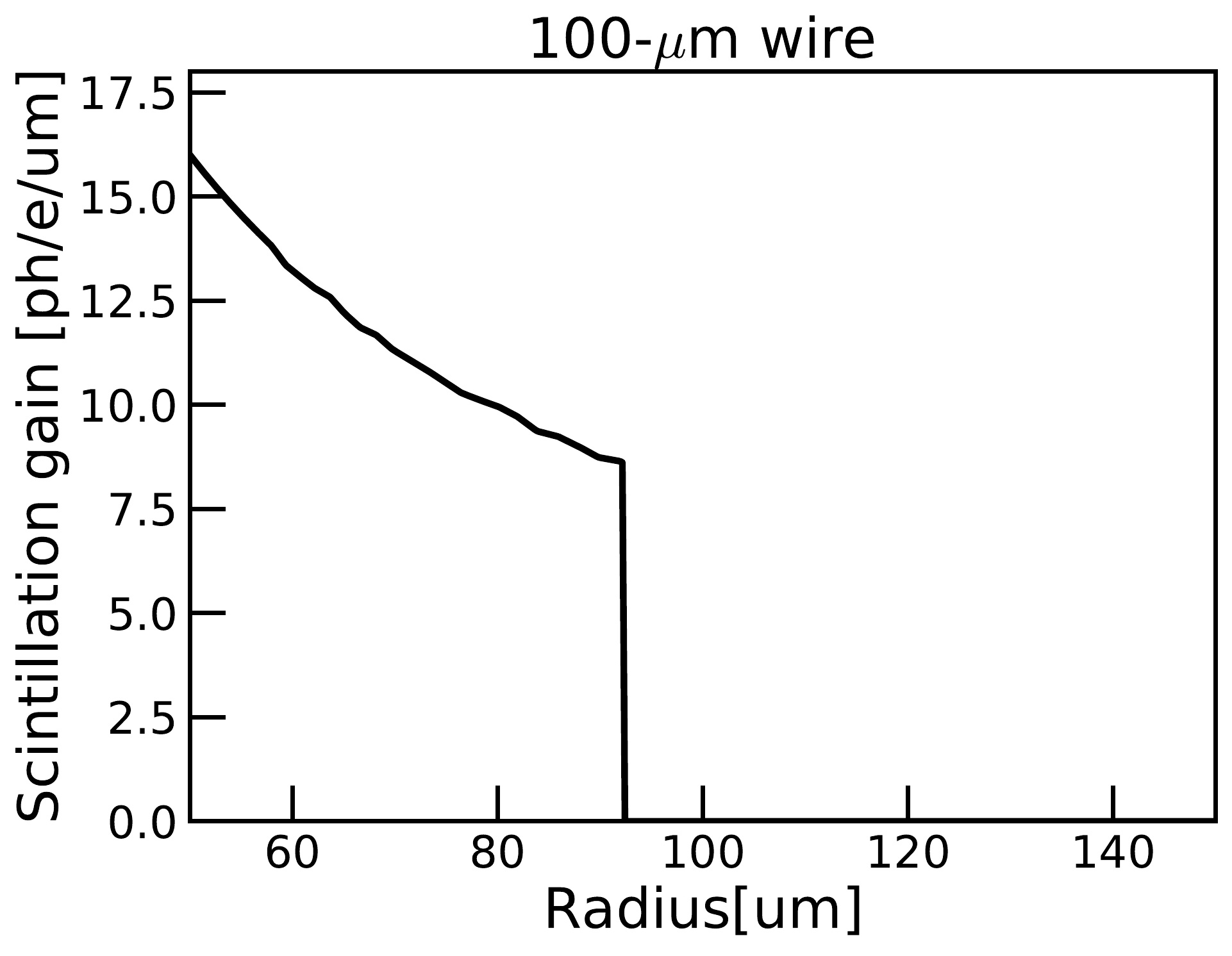}
\caption{
S2 gain as a function of radius for 100-$\mu$m (left) and 200-$\mu$m (right) anode wires.
}
\label{fig:s2gain_vs_radius}
\end{figure}

% results & discussion about simulation results
Fig.~\ref{fig:sim_field_line} shows the simulated local field lines and voltages in gate-anode region for prototype GG-TPC with wire diameters of 100\,$\mu$m and 200\,$\mu$m, respectively.
Fig.~\ref{fig:sim_field_vs_R} shows more quantification of field simulation results at different radial positions for prototype GG-TPC.
With anode voltage at +40\,kV (+28\,kV), GG-TPC with 200-$\mu$m (100-$\mu$m) wires can surpass the threshold field strength of about 412\,kV/cm~\cite{aprile2014measurements} near anode wire for electroluminescence in LXe.
Using the estimated S2 gain factor of (2.09$^{+0.65}_{-0.47}$)$\times$10$^{-2}$\,ph/e$^-$/(kV/cm)/$\mu$m obtained by~\cite{aprile2014measurements}, the amplification factors are calculated to be 792$^{+246}_{-178}$ and 409$^{+127}_{-92}$ ph/e$^-$ (summarized in Table~\ref{tab:amplification_factors} as well), respectively, for GG-TPC with 200-$\mu$m and 100-$\mu$m anode wire.
The proportional scintillation starts at approximately 80\,$\mu$m (41\,$\mu$m) away from the anode wire surface for 200-$\mu$m (100-$\mu$m) wire, which corresponds to an S2 duration of 31\,ns (16\,ns) assuming a saturated drift velocity of $\sim$2.578\,mm/$\mu$s~\cite{aprile2010liquid}.
This indicates that the time profile of the proportional light signals in such GG-TPC is dominated by electron diffusion during drift.
The relation between proportional scintillation gain and distance to anode wire surface for 100-$\mu$m and 200-$\mu$m anode wires can be found in Fig.~\ref{fig:s2gain_vs_radius}.

\begin{figure}[htp]
\centering
\includegraphics[width=0.7\textwidth]{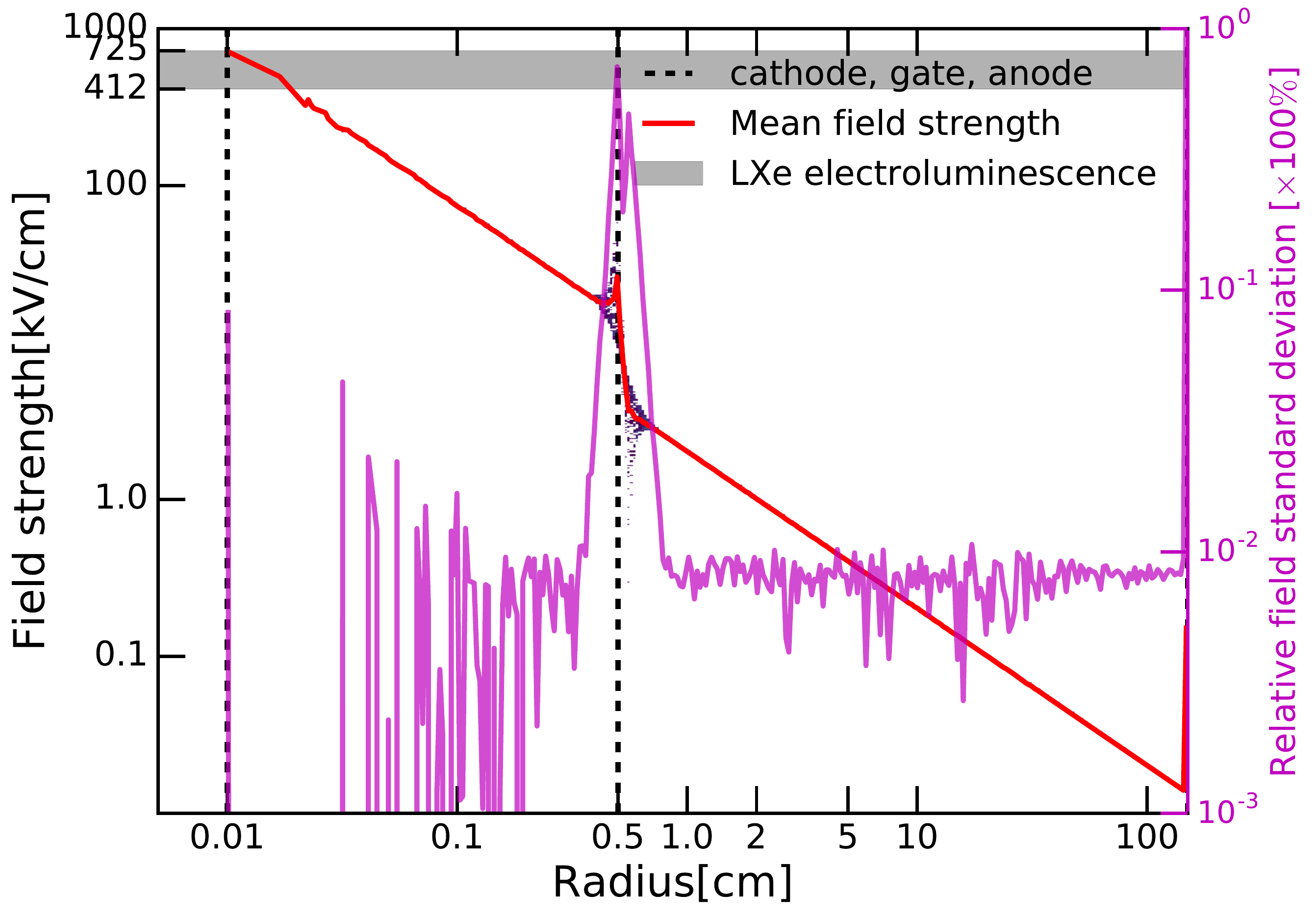}
\caption{
Distribution of TPC volume on field strength vs. radial position for a prototype GG-TPC with radius of 1.5\,m.
The figure description is the same as in Fig.~\ref{fig:sim_field_vs_R}.
}
\label{fig:sim_field_vs_R_largeTPC}
\end{figure}

% Higher anode voltage may reach a field strength above the threshold of avalanche in LXe which is about 720\,kV/cm~\cite{aprile2014measurements}.

A full-scale simulation for prototype GG-TPC is performed with 200-$\mu$m wire, which is shown in Fig.~\ref{fig:sim_field_vs_R_largeTPC}.
%\sout{The anode and gate voltages are increased to +40\,kV and +10\,kV, respectively.}
Results show that the standard deviation of field strength can reach below $\sim$1\% in most part of active volume, except for the regions that are roughly 0.5\,cm apart from the gate and 6\,cm away from cathode in prototype GG-TPC.
Same as traditional dual-phase TPC, the regions near gate and cathode also are to be removed because of the high expected background rate due to radioactivity in wires, from reflector surface, and from outside.
Removal of region that is about 6\,cm away from cathode is compatible with radial cut of tonne-scale detectors, such as XENON1T~\cite{collaboration2018dark}, XENONnT~\cite{aprile2020projected}, PandaX-4T~\cite{zhang2019dark}, and LZ~\cite{akerib2020projected} experiments.
There is no significant extra loss of fiducial volume due to the field variance in GG-TPC.
In active volume with moderate field variance, the field strength ranges from about 40\,V/cm near the cathode to about 3.5\,kV/cm near the gate.
Also the number of gate and cathode wires can be increased in large-scale GG-TPC to provide even lower standard deviation of fields.

\begin{table}[htp]
    \small
    \centering
    \begin{tabular}{c|c|c|c}
    \hline \hline
    Wire diameter [$\mu$m] & Anode voltage [kV] & S2 amplification factor [ph/e$^-$] & Singe e$^-$ S2 duration [ns]\\
    \hline \hline
    100 & 18 & 409$^{+127}_{-92}$  & 16 \\
    200 & 30 & 792$^{+246}_{-178}$ & 31 \\
    \hline \hline
    \end{tabular}
    \caption{
    S2 amplification factors and SE S2 durations for different gate-anode configurations in GG-TPC.
    }
    \label{tab:amplification_factors}
\end{table}

\section{Signal Reconstruction of GG-TPC}
\label{sec:signal_simulation}

The biggest trade-off of GG-TPC is its R-dependent field strength.
Although Section~\ref{sec:field_simulation} shows that the relation between field strength and radial position is well defined, the non-uniform field configuration brings challenges to signal and position reconstructions.
In this section, we study the classification between S1s and single electron S2s, position reconstruction, and ER-NR discriminations in GG-TPC, all of which are important to dark matter direct search.

\subsection{Classification of S1 and SE S2}

Unlike dual phase TPC which has time duration of about several hundred nanoseconds for S2s caused by single electron (SE), GG-TPC has only couple of tens of nanoseconds for SE signals.
Although the large S2 signals still can be distinguished through the signal duration in GG-TPC since electron diffusion dominates the time profile, it has less discrimination power between S1s and SE S2s compared to dual phase TPC.
This may cause the increase of isolated S1 signal rate, thus increase of accidental pileup background.
We would like to argue that discrimination power can be compensated by classification using pulse shape discrimination (PSD).

\begin{figure}[htp]
\centering
\includegraphics[width=0.48\textwidth]{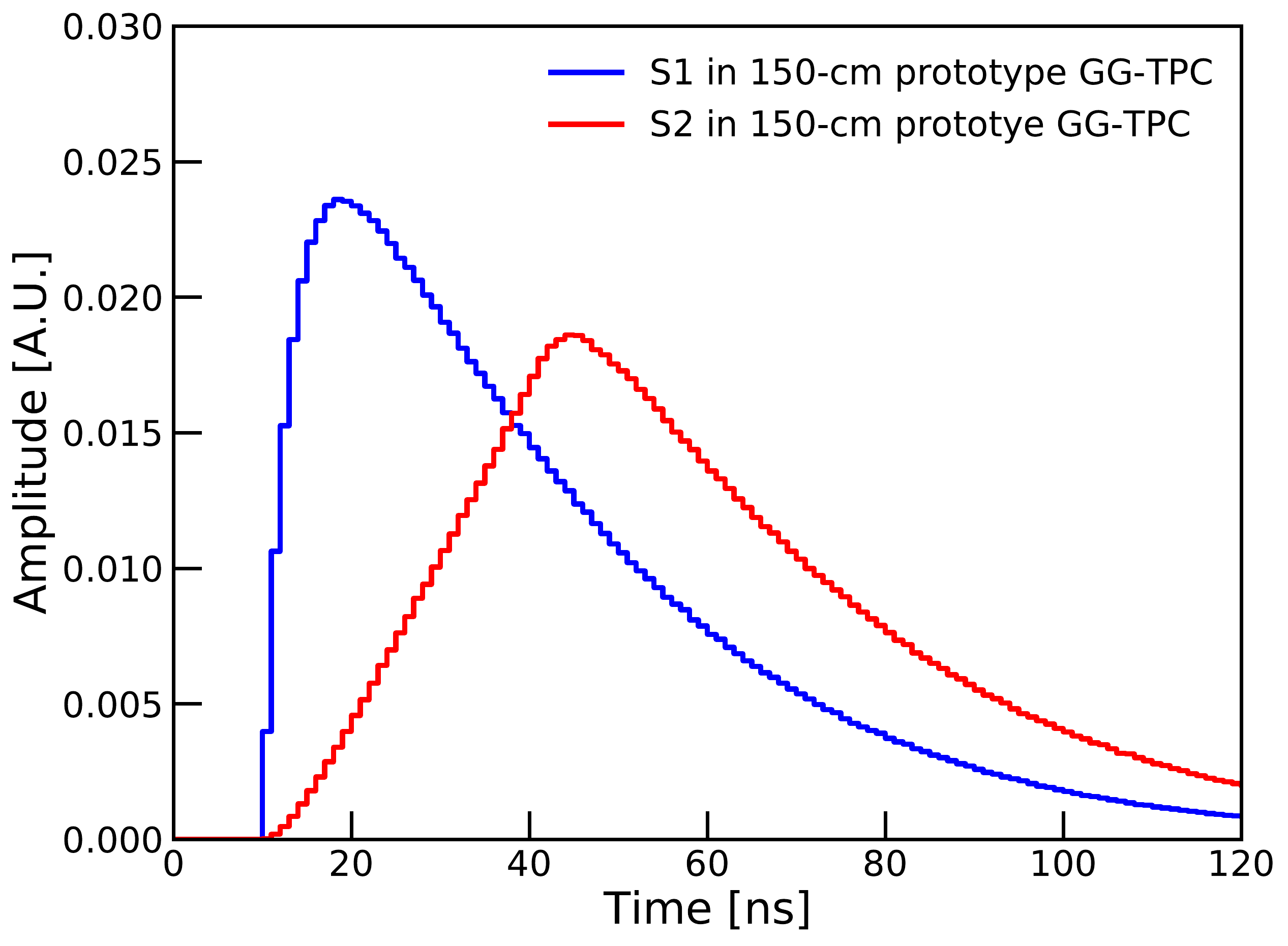}
\includegraphics[width=0.48\textwidth]{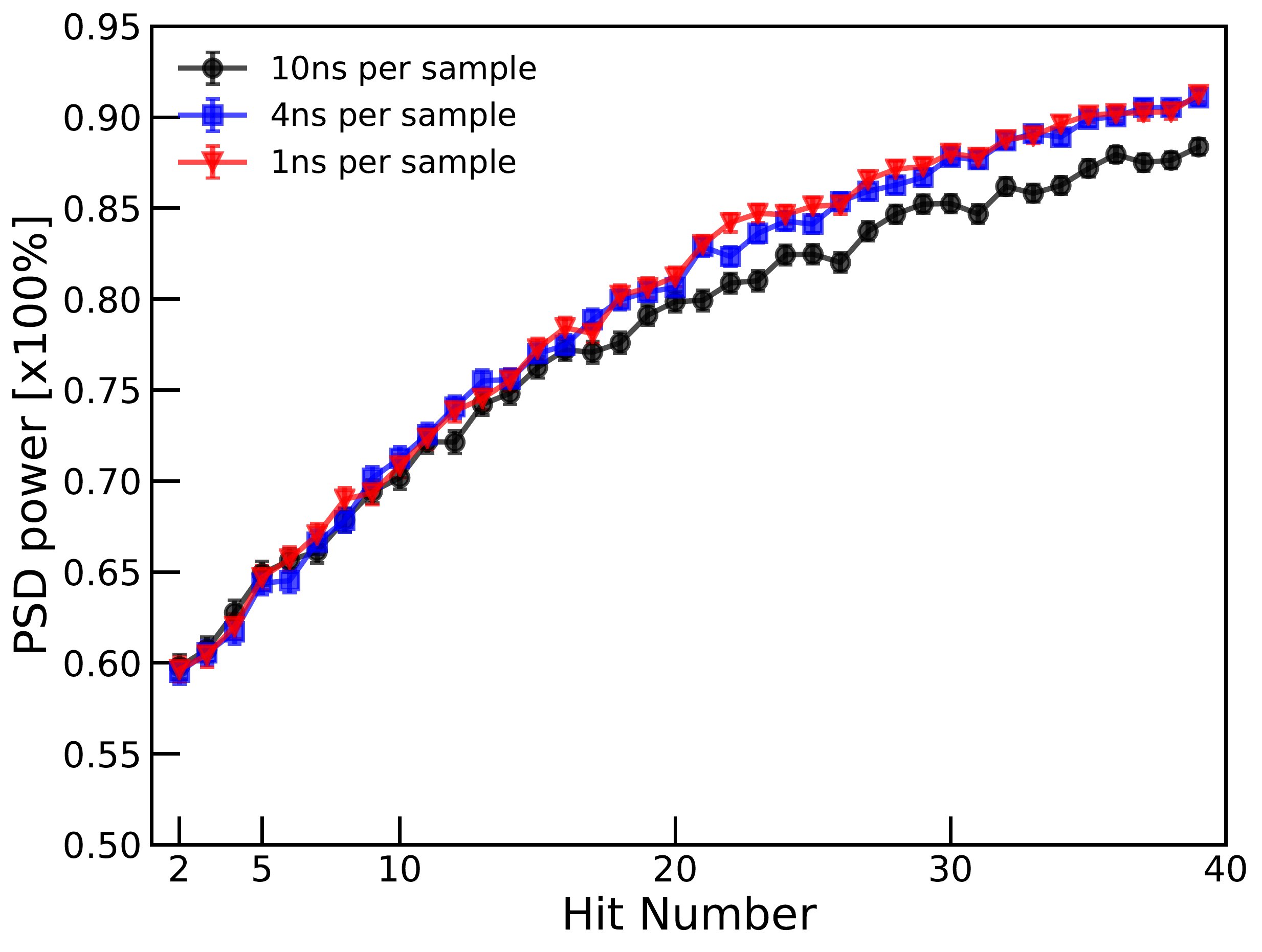}
\caption{
Left plot shows the time profile of S1 and SE S2 in prototype GG-TPC with radius of 1.5\,meter.
Right plot shows the classification accuracy between S1s and SE S2s, using a simple 2-hidden-layer multi-layer perceptron neural network, as a function of hit number.
}
\label{fig:psd}
\end{figure}

% PSD
In LXe, the scintillation light signals are generated through excimer decay~\cite{chepel2013liquid} which has two decay components: fast and slow decay components with lifetime of $\sim$3\,ns and $\sim$24\,ns~\cite{chepel2013liquid} corresponding to singlet and triplet state decays, respectively.
The ratio between the singlet and triplet components depends on the recoil type~\cite{mock2014modeling}, and is about 0.8 for ERs and 7.8 for NRs~\cite{mock2014modeling}.
Therefore, the S1 pulse shape can be modeled as two exponential functions.
On the other hand, SE lights are generated, more or less, uniformly along the accelerated electron trajectory near the anode wire (See Fig.~\ref{fig:s2gain_vs_radius}).
Left plot of Fig.~\ref{fig:psd} shows the expected pulse shape between S1 and SE S2 in prototype GG-TPC.
% \sout{The difference in the pulse shapes between GG-TPC-P1 and GG-TPC-P2 is caused by the optical light propagation in detector.}
The time profile calculation of S1 is based on the singlet-to-triplet ratio of 0.8 for ER, which is more conservative in our analysis compared with 7.8 for NR.
The time profile calculation of S2 is based on the S2 scintillation gain as a function of radius for 200-\,$\mu$m anode wire shown in Fig.~\ref{fig:s2gain_vs_radius}, and on an assumed saturated electron drift velocity of 2.578\,mm/$\mu$s in LXe~\cite{aprile2010liquid}.
Also, ERs are the major background in dark matter direct searches.
The light propagation in GG-TPC is considered and obtained through optical simulation using GEANT4 toolkit~\cite{agostinelli2003geant4}, assuming absorption length of 10\,meter and Rayleigh scattering length of 50\,cm in LXe. 
Light sensor caused pulse shaping of single photon hit is not taken into account, since the shaping depends greatly on type of sensor.
A classification using multi-layer perceptron (MLP) with 2-hidden-layer is conducted to estimate the PSD power for prototype GG-TPC with different sampling sizes (typically 1\,ns, 4\,ns, and 10\,ns per sample).
Training and testing samples are sampled hits based on pulse shapes shown in left plot of Fig.~\ref{fig:psd}.
In the right plot of Fig.~\ref{fig:psd} shows the PSD power (defined as the accuracy of classification) as a function of hit number.
% \sout{GG-TPC-P2 basically has less PSD power over GG-TPC-P1, because of the smearing of pulse shape due to optical light propagation in detector.}
PSD power can reach as high as $\sim$80\% for prototype GG-TPC at hit number of 20, and PSD power is reduced to about 60\% when hit number gets as low as 2.
This indicates that reaching high charge amplification factor and high light collection efficiency for proportional scintillation will be important in terms of achieving high classification accuracy between S1s and SE S2s in GG-TPC.
% In addition, the sample size also plays a role in the PSD power for GG-TPC-P1 (improved by approximately 5\% when reducing sample size from 10\,ns to 4\,ns or 1\,ns), but makes little difference for GG-TPC-P2.
In addition, the sample size plays a minor role in PSD power for the prototype GG-TPC.

\subsection{Position Reconstruction}

% importance of position reconstruction
3-D position reconstruction is one of the most critical features of TPC. 
It allows TPC to be capable of rejecting external gamma-rays and surface backgrounds by fiducialization, discriminating against neutrons by identifying multiple scatters, and reaching high energy resolution by correcting the signals for their spatial dependence.
Usually, one spatial coordinate in a TPC is reconstructed through the time difference between scintillation and ionization signals (drift time) which has relatively high time resolution, and the rest two coordinates are reconstructed using the signal pattern on instrumented sensors (PMTs in dual phase LXe TPCs~\cite{aprile2017xenon1t, cao2014pandax, akerib2015lux} or wire planes in liquid argon TPCs~\cite{amerio2004design, soderberg2009microboone}) which is limited by the granularity of sensors.
GG-TPC has its radial coordinate reconstructed using drift time, but it does not have good angular reconstruction resolution because the proportional scintillation occurs in the center axis of the TPC.
However as mentioned in previous section, angular distribution of events is less important than radial and zenith distribution for a GG-TPC in dark matter direct search since it provides minor information for rejecting backgrounds and correcting signals in an axial symmetrical detector.
The zenith position is, on the other hand, critical for rejecting the backgrounds from two bases of the cylindrical volume (we name the two bases as top and bottom for convenience in later text).

\begin{figure}[htp]
    \centering
    % \includegraphics[width=0.48\textwidth]{plots/s1_pattern_10cm.pdf}
    % \includegraphics[width=0.48\textwidth]{plots/s2_pattern_10cm.pdf}
    % \hfill
    \includegraphics[width=0.48\textwidth]{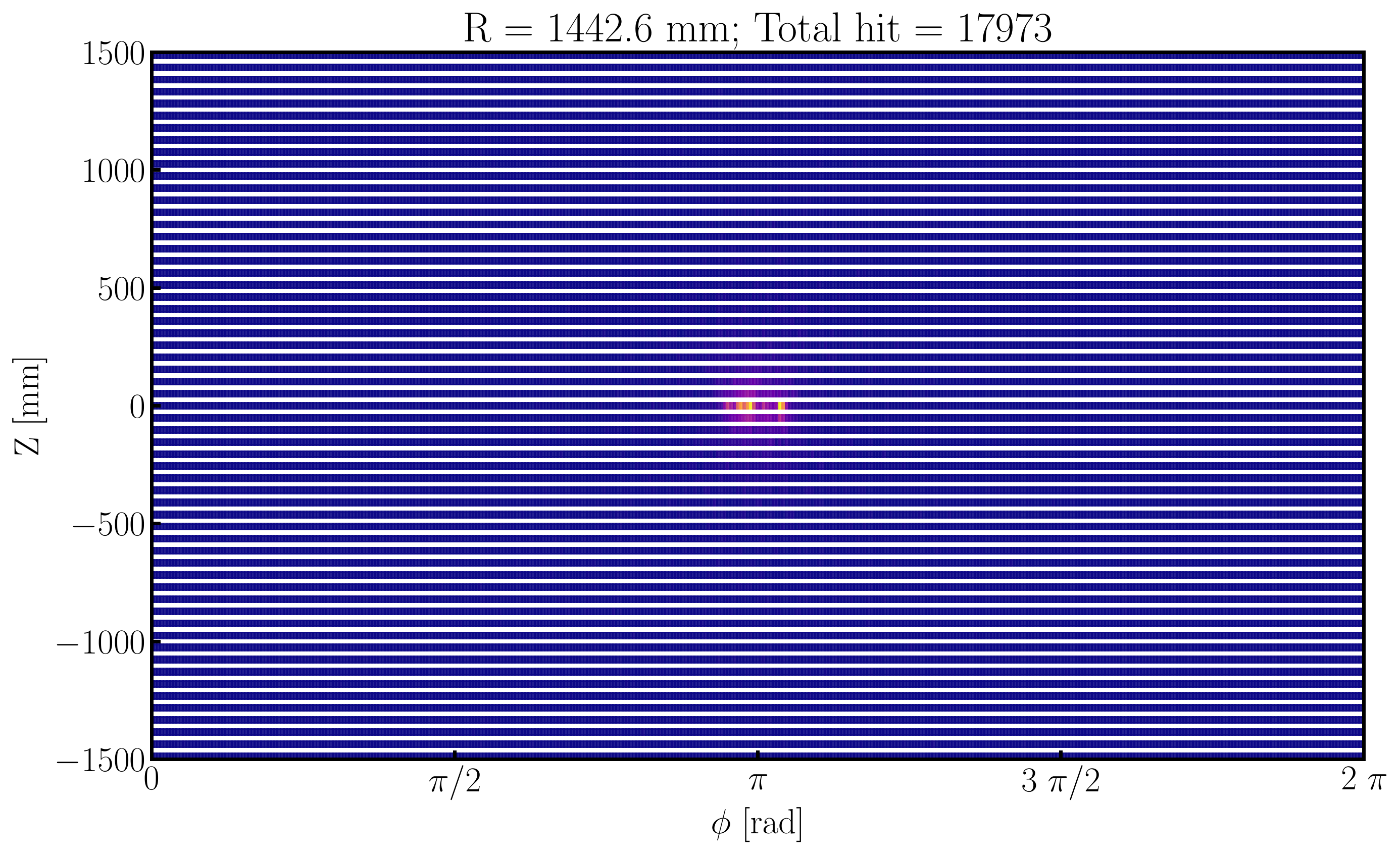}
    \includegraphics[width=0.48\textwidth]{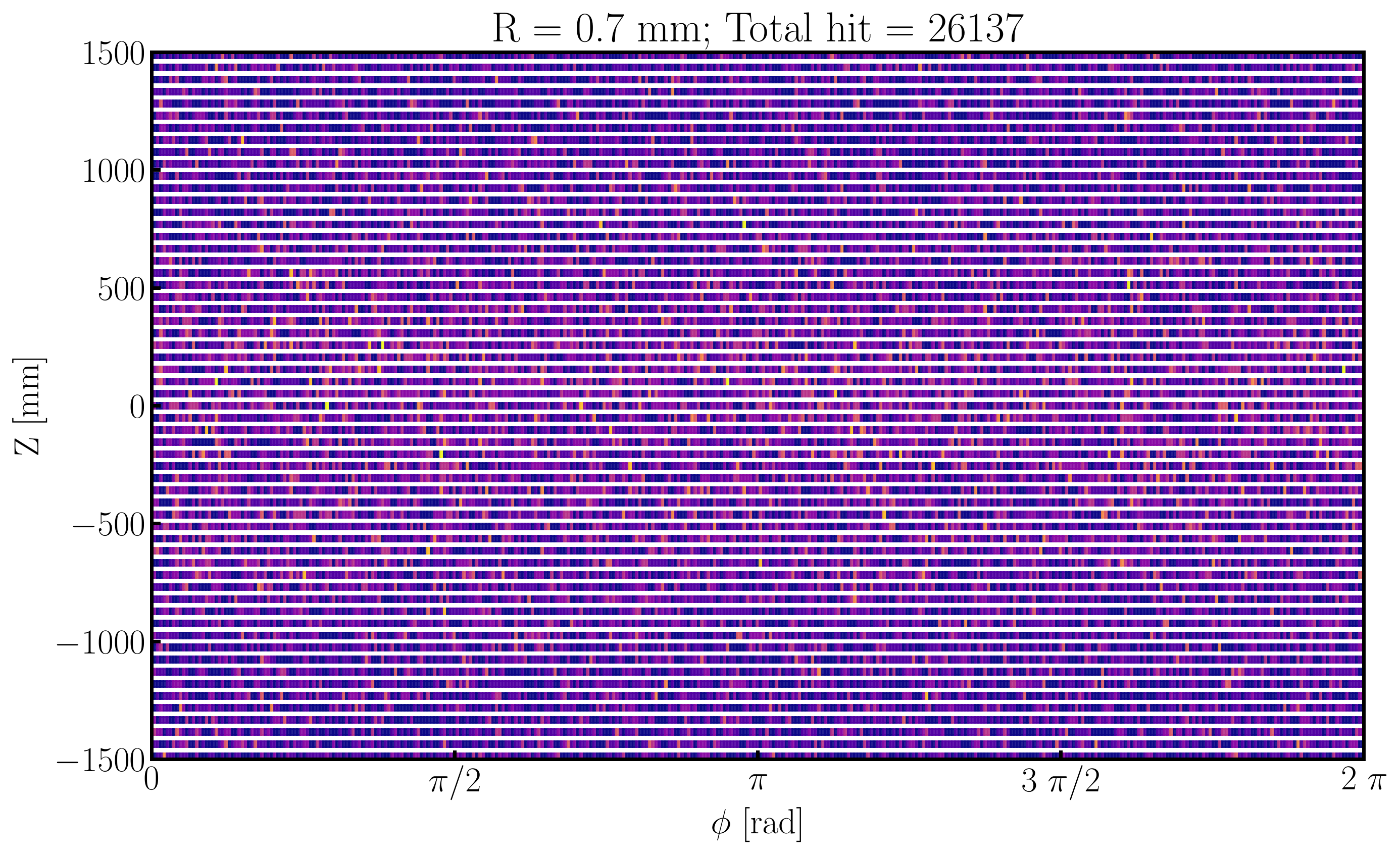}
    \caption{
    Hit pattern of two sample simulated events.
    % The top and bottom rows are events from GG-TPC-P1 and GG-TPC-P2, respectively.
    The left and right plots are events that are close to side wall and central anode, respectively.
    The radial position and total hit number of each event are given in the title above each plot.
    }
    \label{fig:pattern}
\end{figure}

\begin{figure}[htp]
    \centering
    \includegraphics[width=0.7\textwidth]{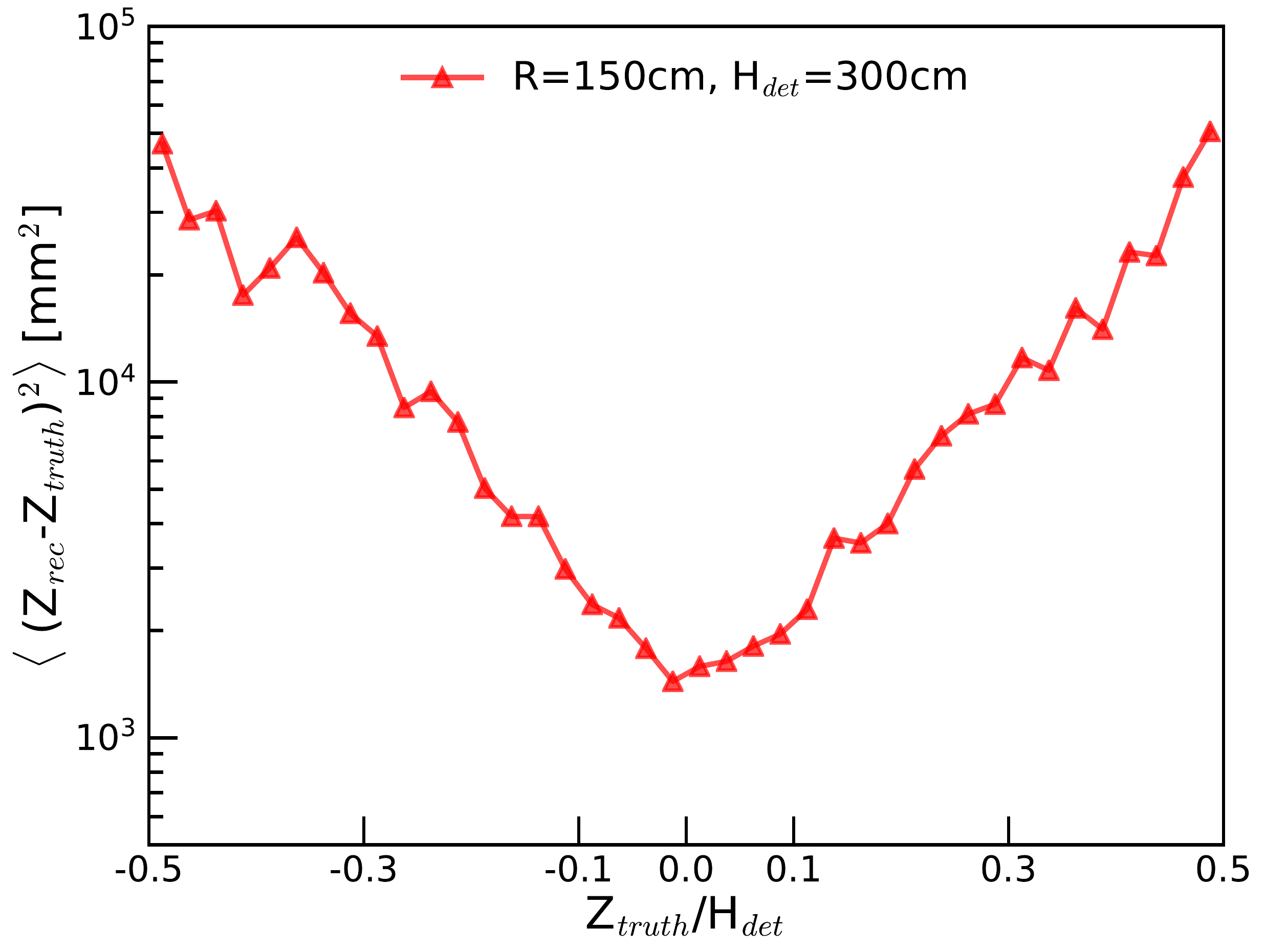}
    \caption{
    PMT pattern reconstructed $Z$ resolution as a function of $Z$ normalized by detector height.
    % The blue circles and red triangles represent the results from GG-TPC-P1 and GG-TPC-P2, respectively.
    }
    \label{fig:posrec_wallpmt_only}
\end{figure}

% Reconstruction using wall pmts & description of optical simulation
In a GG-TPC shown in Fig.~\ref{fig:single_phase_tpc}, the hit pattern of the proportional scintillation on PMTs can be used to reconstruct the zenith position.
The reconstruction is performed using a Convolutional Neural Network (CNN) based algorithm which is trained on data from optical simulation done using GEANT4 toolkit~\cite{agostinelli2003geant4}, assuming absorption length of 10\,meter and Rayleigh scattering length of 50\,cm in LXe.
CNN has depth (number of layers) of 7 for the prototype GS-TPC.
Each depth includes a convolutional layer with kernel size of 3 and a max-pooling.
Four hundred thousand simulated events are used as training sample, and one hundred thousand are used as testing sample.
In simulated events, the number of hits is about 1\,k to 2\,k.
The optical simulation assumes an absorption length of 10\,m and Rayleigh scattering length of 50\,cm in LXe.
The reflectivity of teflon used is assumed to be 99\%.
Teflon planes are placed at top and bottom of GG-TPC to increase light collection.
The PMT window size is 1-inch square.
% Simulations are performed for both GG-TPC-P1 and GG-TPC-P2.
Numbers of PMT rings are instrumented on the side of GG-TPC.
PMT rings are spaced with a fixed distance on $Z$ axis to have a light coverage of 50\% on the side wall.
There are totally 21948 PMTs for prototype GG-TPC.
The PMT arrangements and hit patterns of four sample signals are shown in Fig.~\ref{fig:pattern}.
The two sample hit patterns are with events all occur on $Y$ axis and with $Z$=0\,mm.
One of them is very close to TPC side, and one of them near the center axis.
Visually, the pattern is more spread as the vertex is further away from the cylindrical side wall.
Fig.~\ref{fig:posrec_wallpmt_only} gives the reconstructed $Z$ resolution as a function of $Z$ position of events.
% For GG-TPC-P1, the $Z$ resolution can reach approximately 2\,cm at the top and bottom， and minimally about 3\,mm in the center of TPC.
The $Z$ resolution is not optimal because Rayleigh scatter smears out $Z$ information of the original event.
The $Z$ resolution is  $\sim$20\,cm at top and bottom, and $\sim$4.5\,cm in the center.

% Reconstruction using Top-Bottom asymmetry

\begin{figure}[htp]
    \centering
    \includegraphics[width=0.5\textwidth]{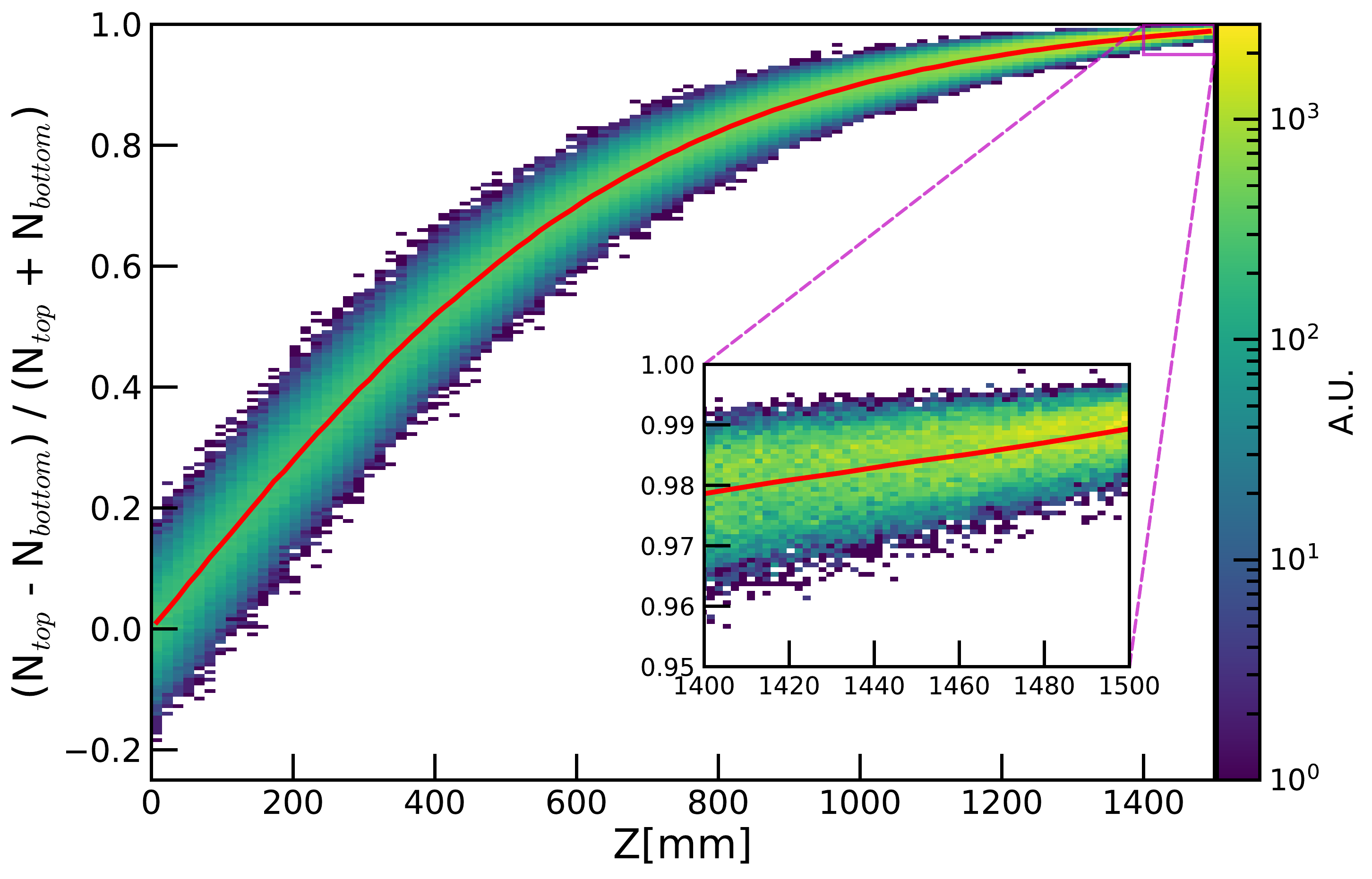}
    \includegraphics[width=0.48\textwidth]{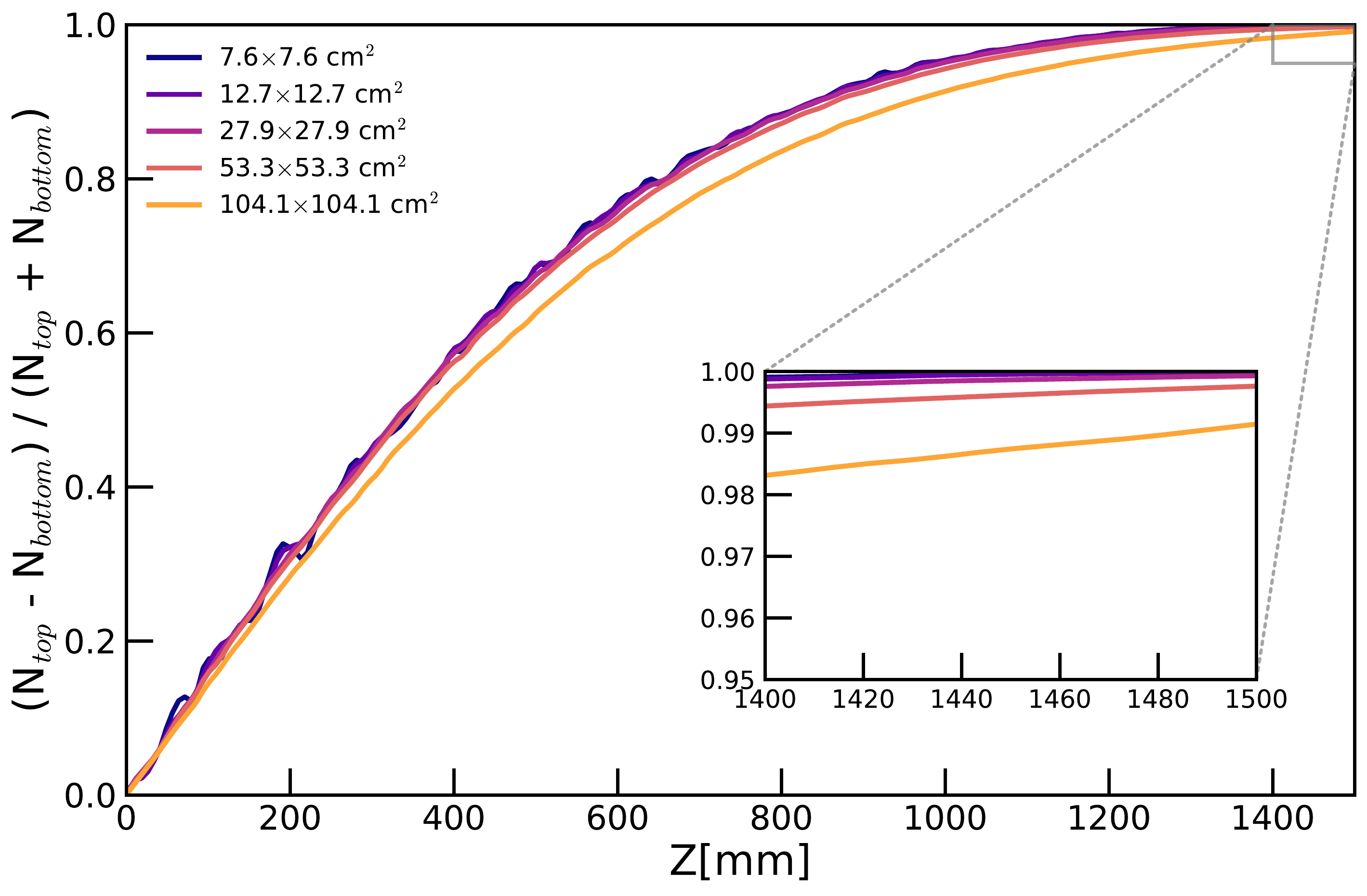}
    \caption{
    Left plot shows TBA distribution of simulated events in prototype GG-TPC with radius of 150\,cm and height of 300\,cm. 
    1681 PMTs (41$\times$41) are placed at both top and bottom.
    Right plot shows the expected TBA as a function of Z for different scenarios of number of kept top/bottom PMTs.
    }
    \label{fig:tba_distribution}
\end{figure}

\begin{figure}[htp]
    \centering
    \includegraphics[width=0.9\textwidth]{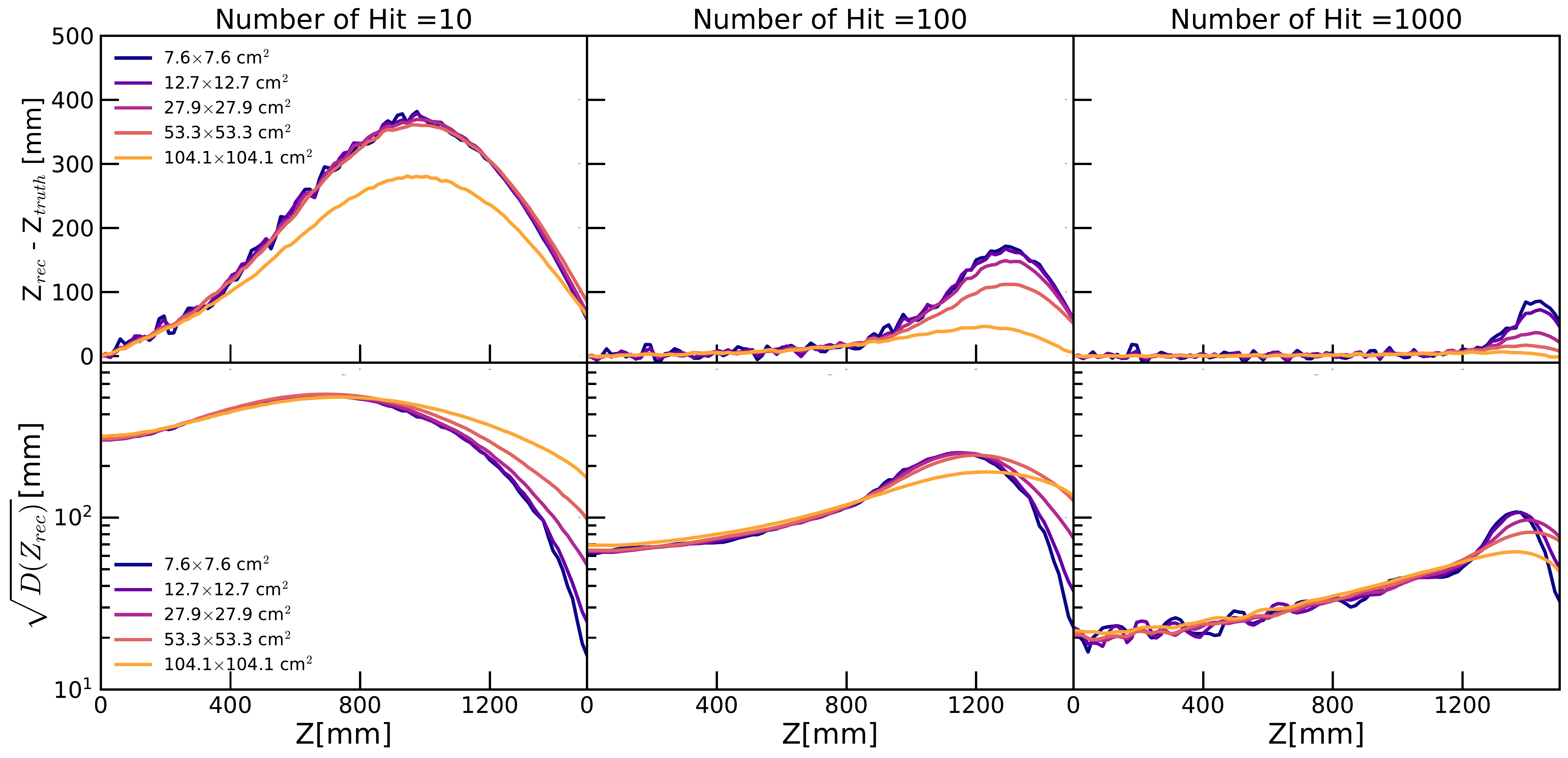}
    \caption{
    Top and bottom rows show the bias and resolution of TBA-based reconstruction, respectively, for different scenarios of number of kept PMTs.
    The left, middle, and right columns give the bias and resolution with number of hits of 10, 100, and 1000, respectively.
    }
    \label{fig:tba_discrimination_vs_z}
\end{figure}

\begin{figure}[htp]
    \centering
    \includegraphics[width=0.48\textwidth]{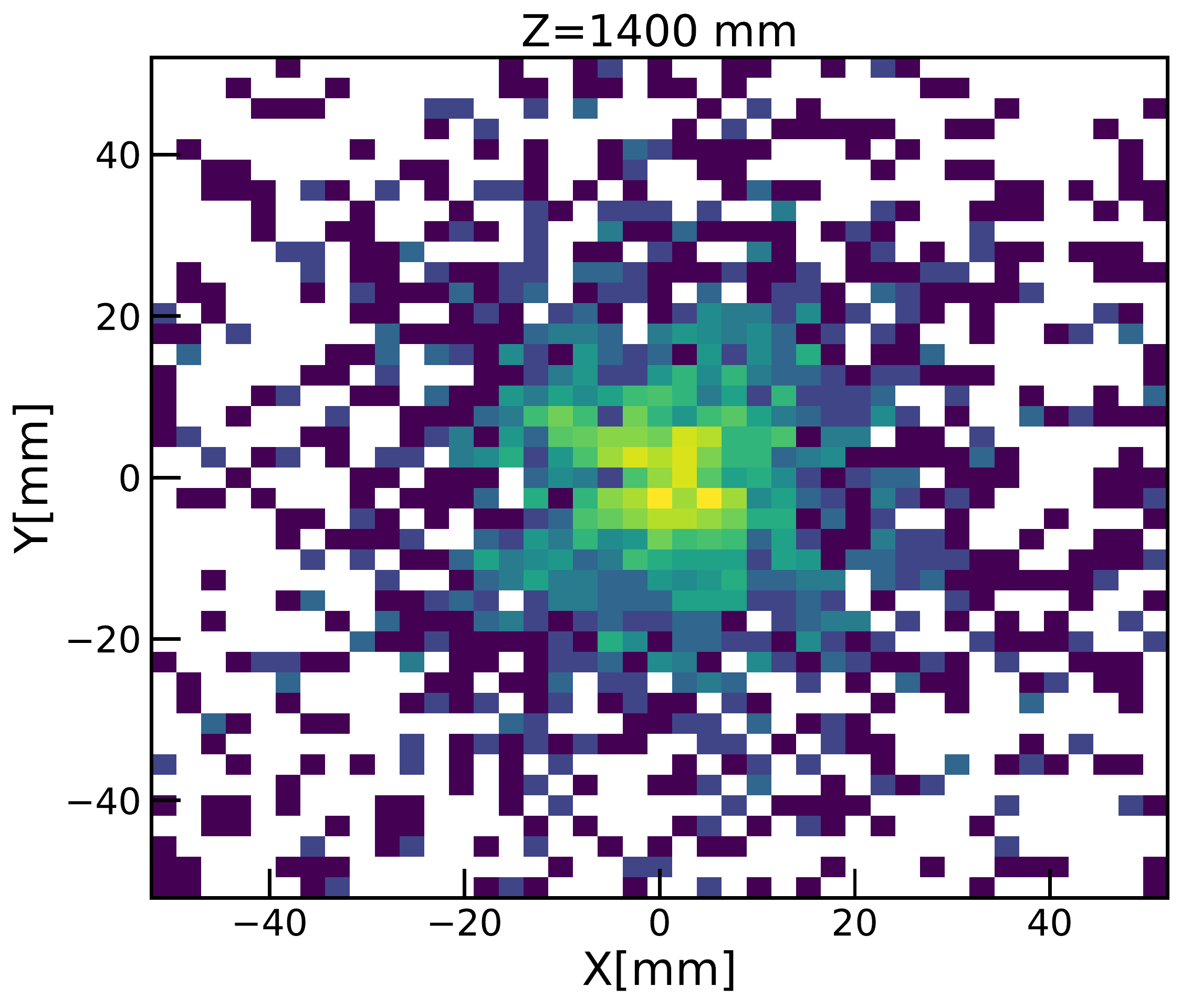}
    \includegraphics[width=0.48\textwidth]{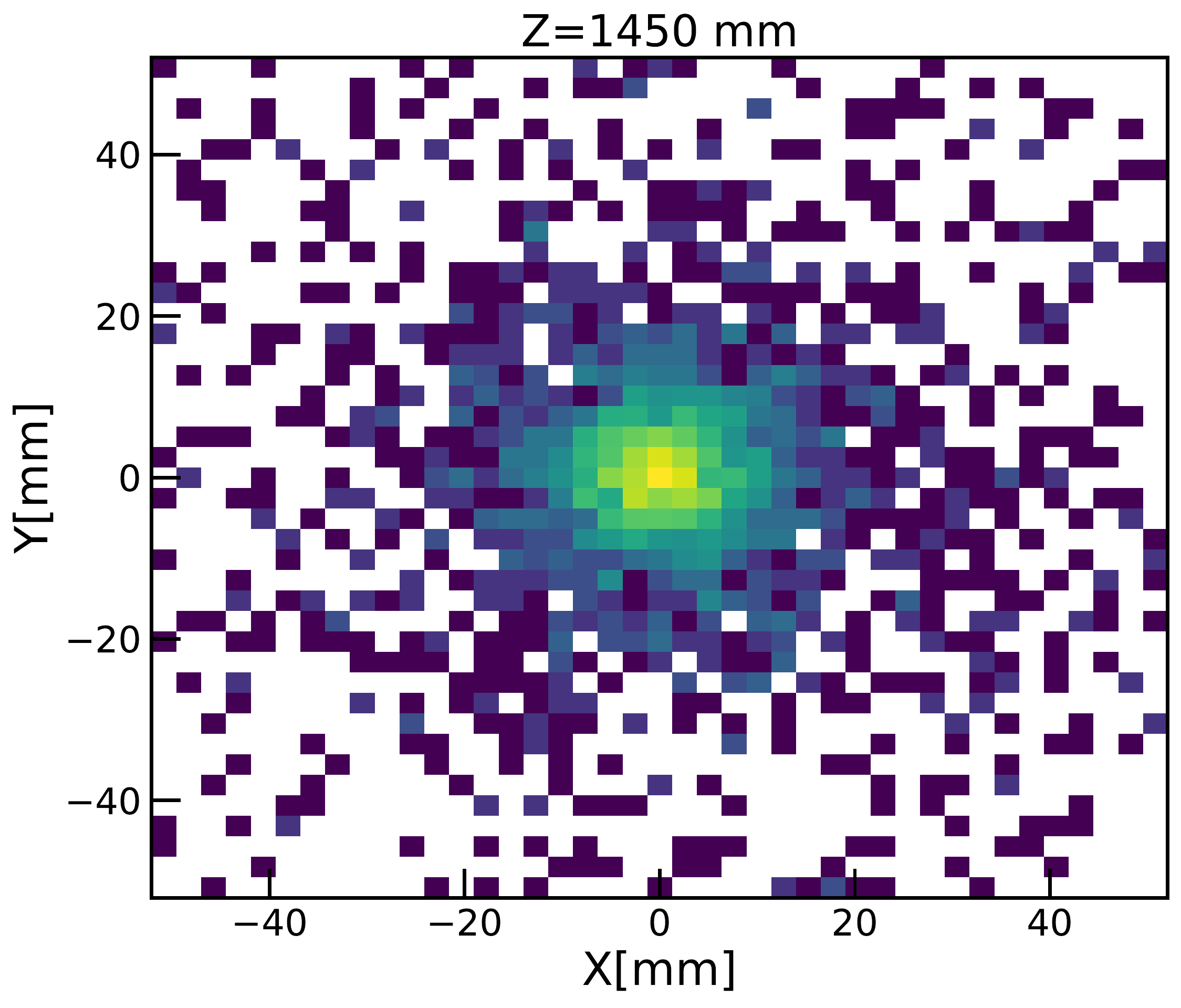}
    \includegraphics[width=0.48\textwidth]{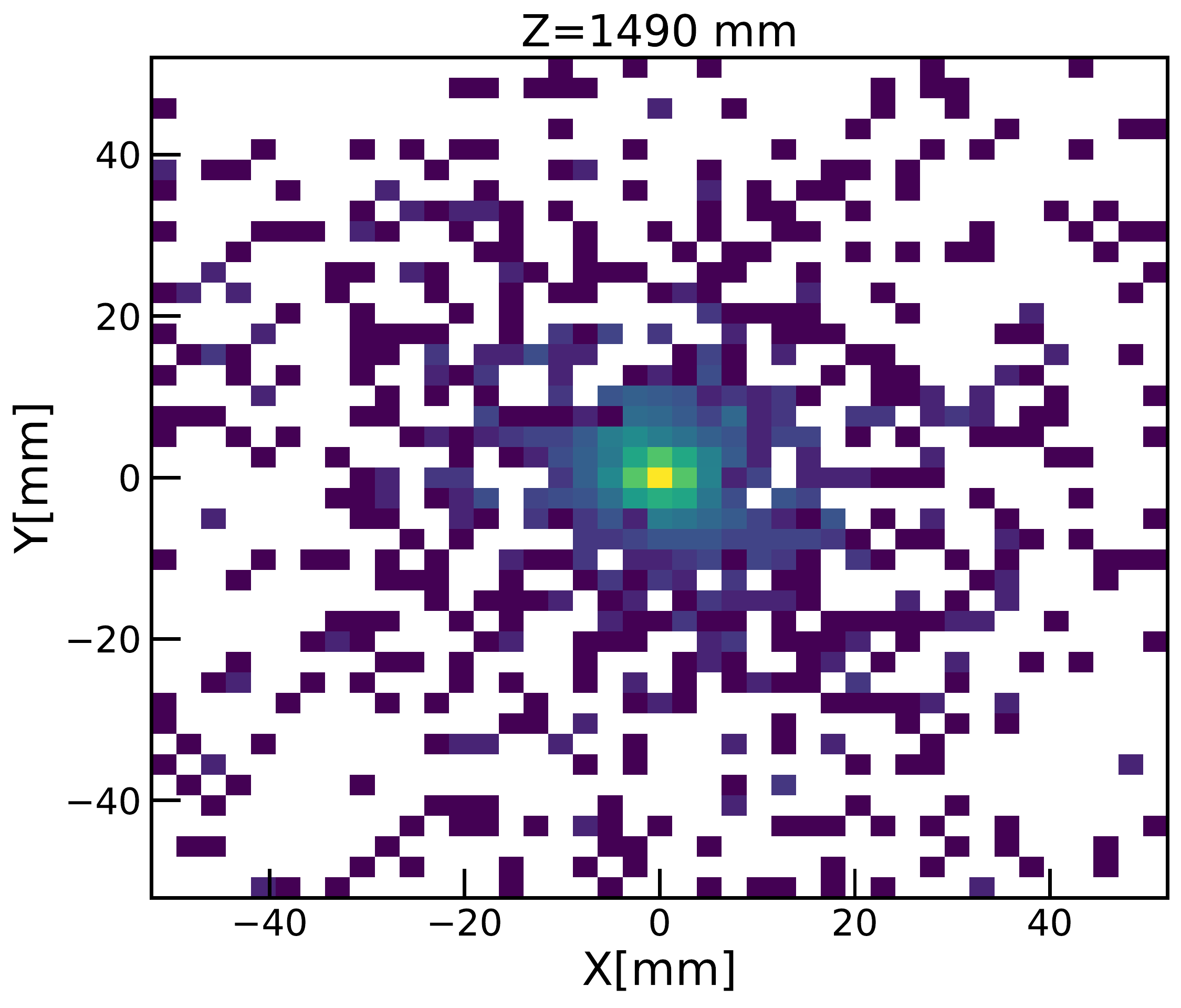}
    \includegraphics[width=0.48\textwidth]{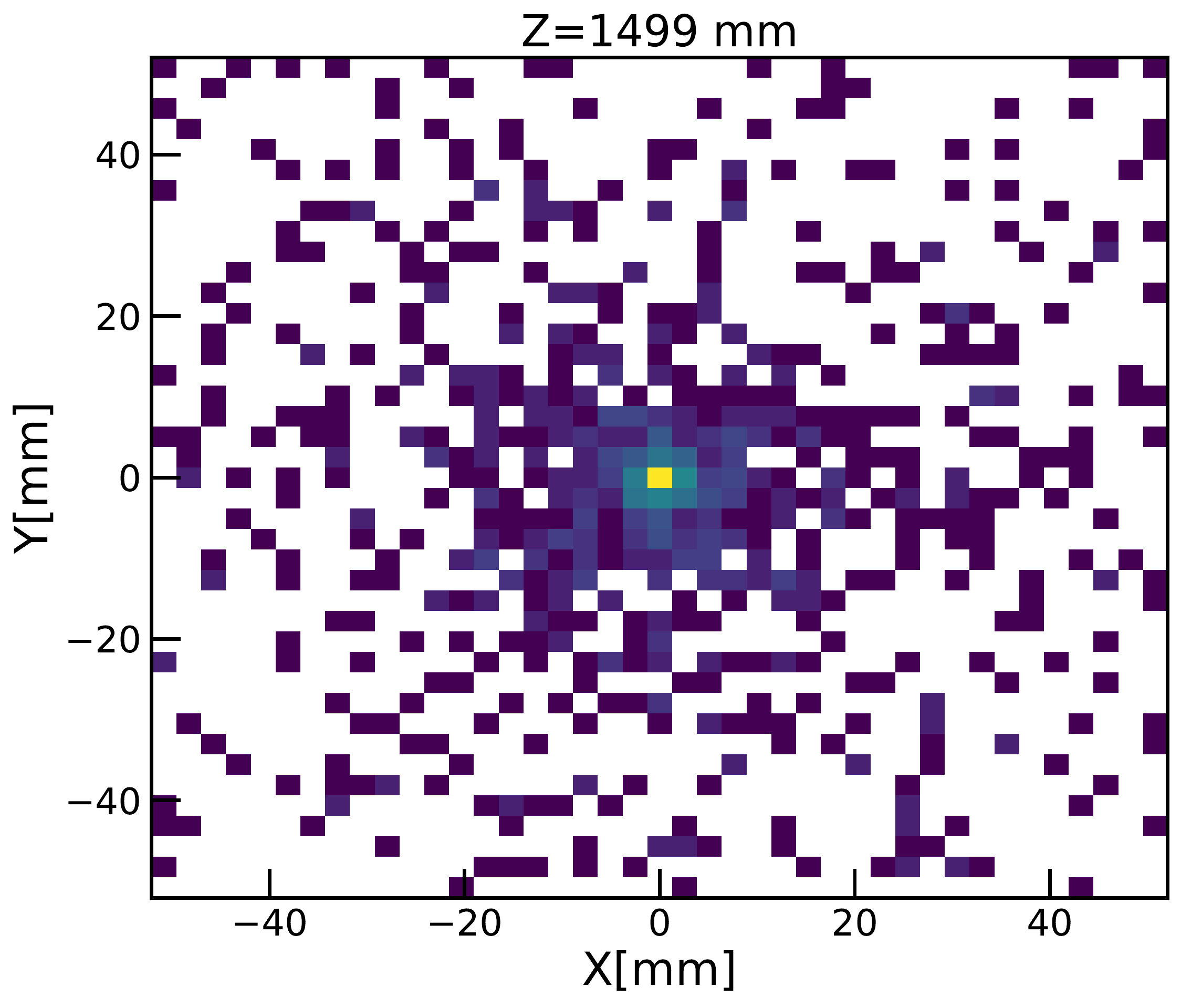}
    \caption{
    Hit patterns on top 41$\times$41 PMTs of simulated events.
    Four events happen 100, 50, 10, and 1\,mm away from top plane, respectively.
    }
    \label{fig:top_pattern}
\end{figure}

The $Z$ resolution near top and bottom edges is critical and 20\,cm is not an acceptable resolution for future 3rd generation detector.
One possible way to improve this is to have light sensors also placed at the top and bottom.
Top-bottom asymmetry (TBA) of S2 and S2 patterns on top/bottom PMTs can provide extra reconstruction power of $Z$ position.
We perform an additional simulation of prototype GG-TPC with PMTs placed at both top and bottom surfaces.
% Each matrix covers roughly a 1\,m$\times$1\,m area.
The $Z$ positions of simulated events are sampled uniformly from -150\,cm to 150\,cm (with 0 as the center of GG-TPC).
In Fig.~\ref{fig:tba_distribution}, the left plot shows the TBA distribution of 1 million simulated events and mean TBA as a function of $Z$ position with 41$\times$41 PMTs placed at top/bottom plane, which can be used to reconstructed $Z$.
It is worth noting that placing light sensors at top and bottom of GG-TPC may bring challenges in maintaining good field uniformity in top and bottom regions, and may increase radioactive background (gamma-rays and neutrons) from PMT material.
It is preferrable to minimize the number of PMTs at top and bottom.
In this work, we check the reconstruction results with different number of 1-inch PMTs placed in the central area of the top and bottom basis plane: 3$\times$3, 5$\times$5, 11$\times$11, 21$\times$21, and 41$\times$41 PMTs, corresponding to area coverage of 7.6$\times$7.6, 12.7$\times$12.7, 27.9$\times$27.9, 53.3$\times$53.3, and 104.1$\times$104.1\,cm$^2$.
Right plot of Fig.~\ref{fig:tba_distribution} gives the expected TBA (with sufficiently large number of hits on top and bottom PMTs) as a function of $Z$ position, for different scenarios of number of top/bottom PMTs as mentioned above.
% Using simple toy MC, we can derive the resolution of TBA-based position reconstruction.
Fig.~\ref{fig:tba_discrimination_vs_z} shows the bias and resolution of TBA-based reconstruction as a function of $Z$ with number of hits on top/bottom PMTs of 10, 100 and 1000, respectively.
The reconstruction resolution $\sqrt{D(Z_{rec})}$ is defined as the standard deviation of reconstructed $Z$.
The left column of Fig.~\ref{fig:tba_discrimination_vs_z} shows that the TBA-based reconstruction has large bias due to very low statistics of number of hits, as well as $>$10\,cm resolution.
Most importantly due to the low light collection efficiency of the opposite PMTs, the TBAs of events near top/bottom basis surface are most likely to be 1, leading to large bias and reduced reconstruction power near top/bottom surface, the region of which is of the most interest in terms of background rejection through fiducialization.
Despite the regions near top and bottom edges, the reconstruction performances between different scenarios of number of kept PMTs have minor difference.

\begin{figure}[htp]
    \centering
    \includegraphics[width=0.7\textwidth]{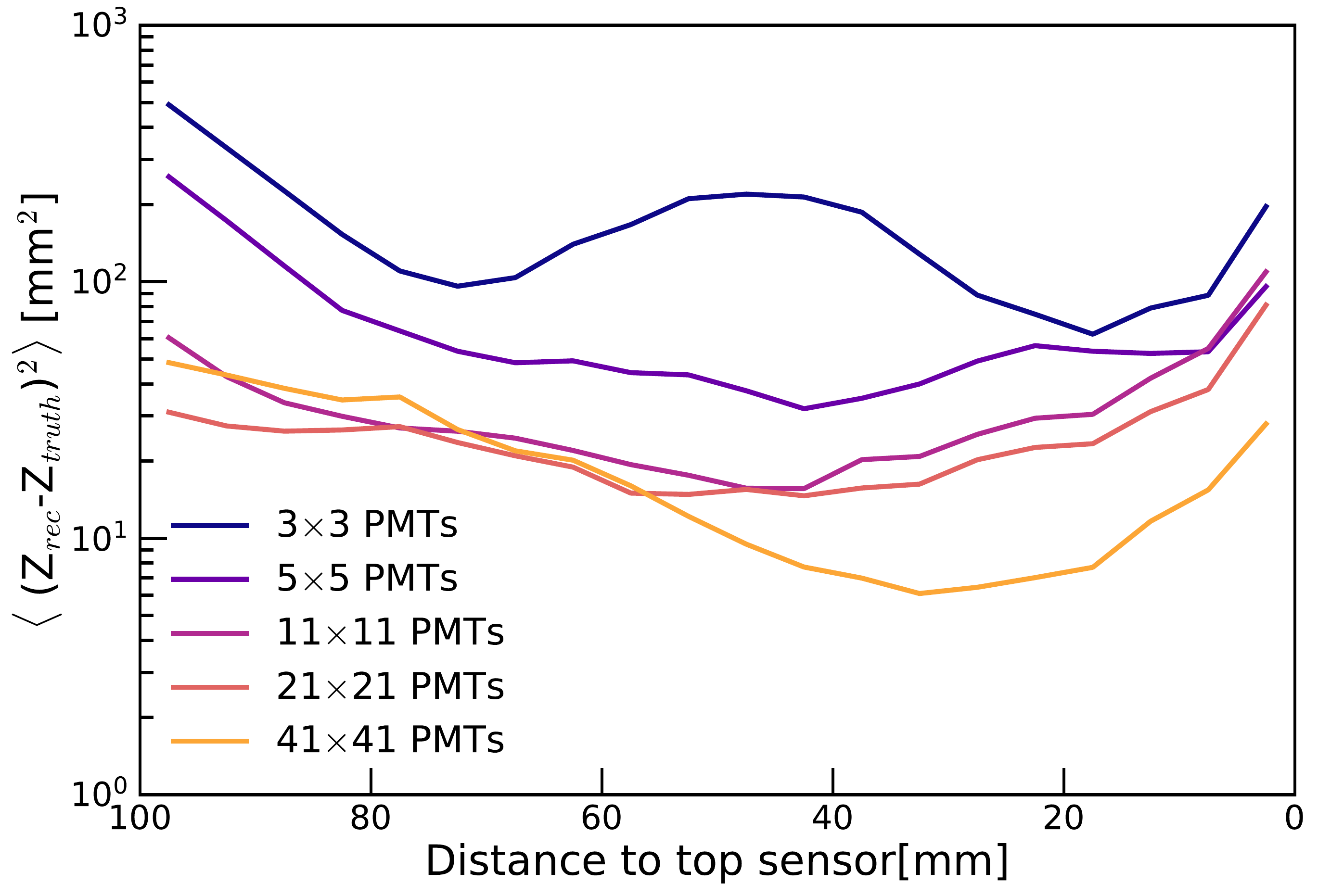}
    \caption{
    Reconstruction resolution as a function of distance to top plane for different scenarios of kept number of PMTs.
    }
    \label{fig:pattern_z_discrimination}
\end{figure}

% Reconstruction further using pattern on top PMTs
Besides TBA, the hit pattern on top/bottom PMTs can be used to conduct $Z$ reconstruction, especially in near top/bottom regions because of the proximity of top/bottom PMTs to proportional lights.
Fig.~\ref{fig:top_pattern} shows the normalized hit pattern on top PMTs (41$\times$41) of four events with different distances to top plane.
Number of hits for these events and later samples in CNN reconstruction for training and testing are roughly between 700 to 3000, depending on number of kept PMTs.
It is visible that the hit pattern gets more spread as the event happens closer to top.
We perform CNN-based reconstruction for different scenarios of kept number of PMTs. 
The CNN architecture has 1, 1, 2, 3, 4 layers for 3$\times$3, 5$\times$5, 11$\times$11, 21$\times$21, and 41$\times$41 kept PMTs, respectively.
Each layer includes a convolutional layer with kernel size of 3 and a max-pooling.
There is minor difference in the performance of the reconstruction for different scenarios.
The slight difference between different scenarios can be due to the different CNN architecture (depth and complexity), which is out of scope of this work and can be further optimised with realistic detector.
The results are shown in Fig.~\ref{fig:pattern_z_discrimination}.
In the region that is within 10\,cm away from top/bottom plane, the resolution of $Z$ that is reconstructed using hit pattern on top/bottom PMTs can reach sub-cm level when hit number is at about one thousand.
In addition, it does not need a large number of PMTs covering top/bottom plane to reach a high $Z$ resolution.
The nine PMTs in the very center of top/bottom plane have the most pattern information for reconstruction.

\subsection{Particle Discrimination}

% illustrate the difficulty
Discrimination between particles inducing ERs and NRs is available in LXe using S2-S1 ratio.
However, the light and charge yields of an energy deposition are related to the electric field strength.
Different from the traditional single or dual TPC with electrode planes having uniform field strength in active volume, GG-TPC intrinsically has electric field strength depending on its radial position.
Therefore simply selecting events from fiducial volume in GG-TPC will result in bad discrimination power since events from low- and high-field regions are overlapping with each other.
We would like to point out that GG-TPC can retain compatible discrimination power compared to a traditional dual phase TPC, by using information of radial position when discriminating particles.
Because GG-TPC has excellent position reconstruction resolution on radial axis, we would argue the loss of discrimination power in GG-TPC compared to a traditional dual phase TPC is at minimum.

\begin{figure}[htp]
    \centering
    \includegraphics[width=0.9\textwidth]{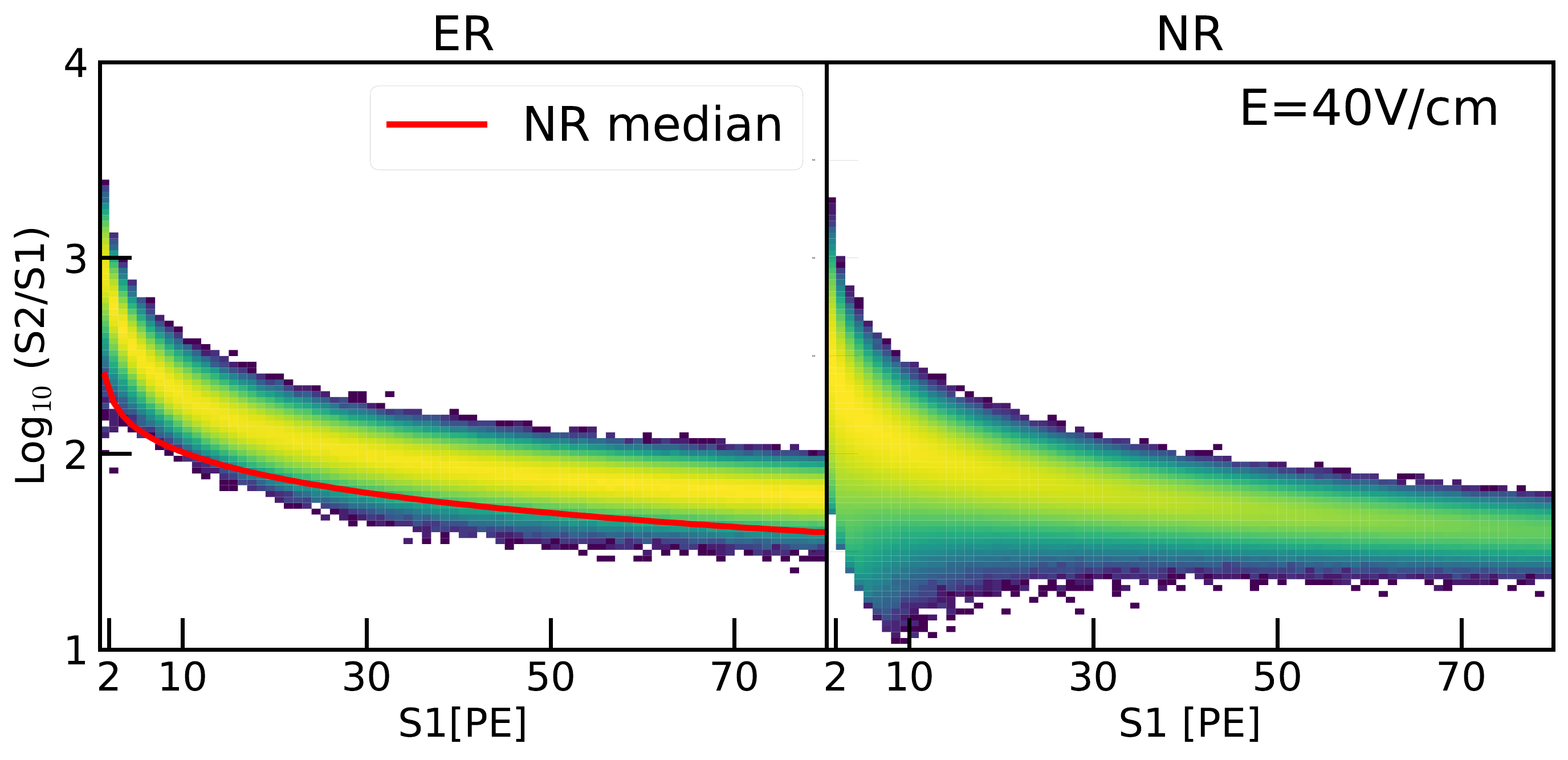}
    \includegraphics[width=0.9\textwidth]{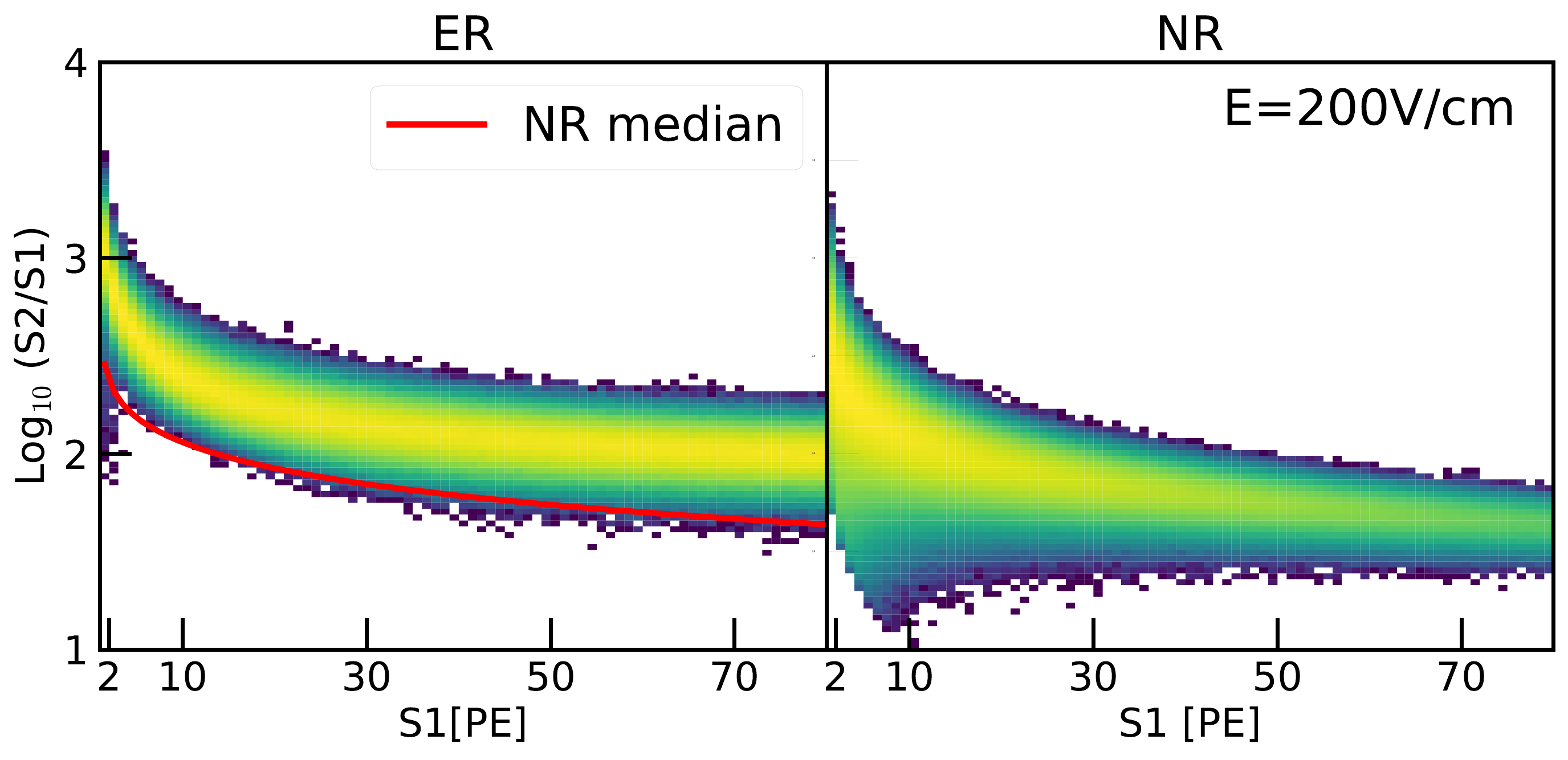}
    \includegraphics[width=0.9\textwidth]{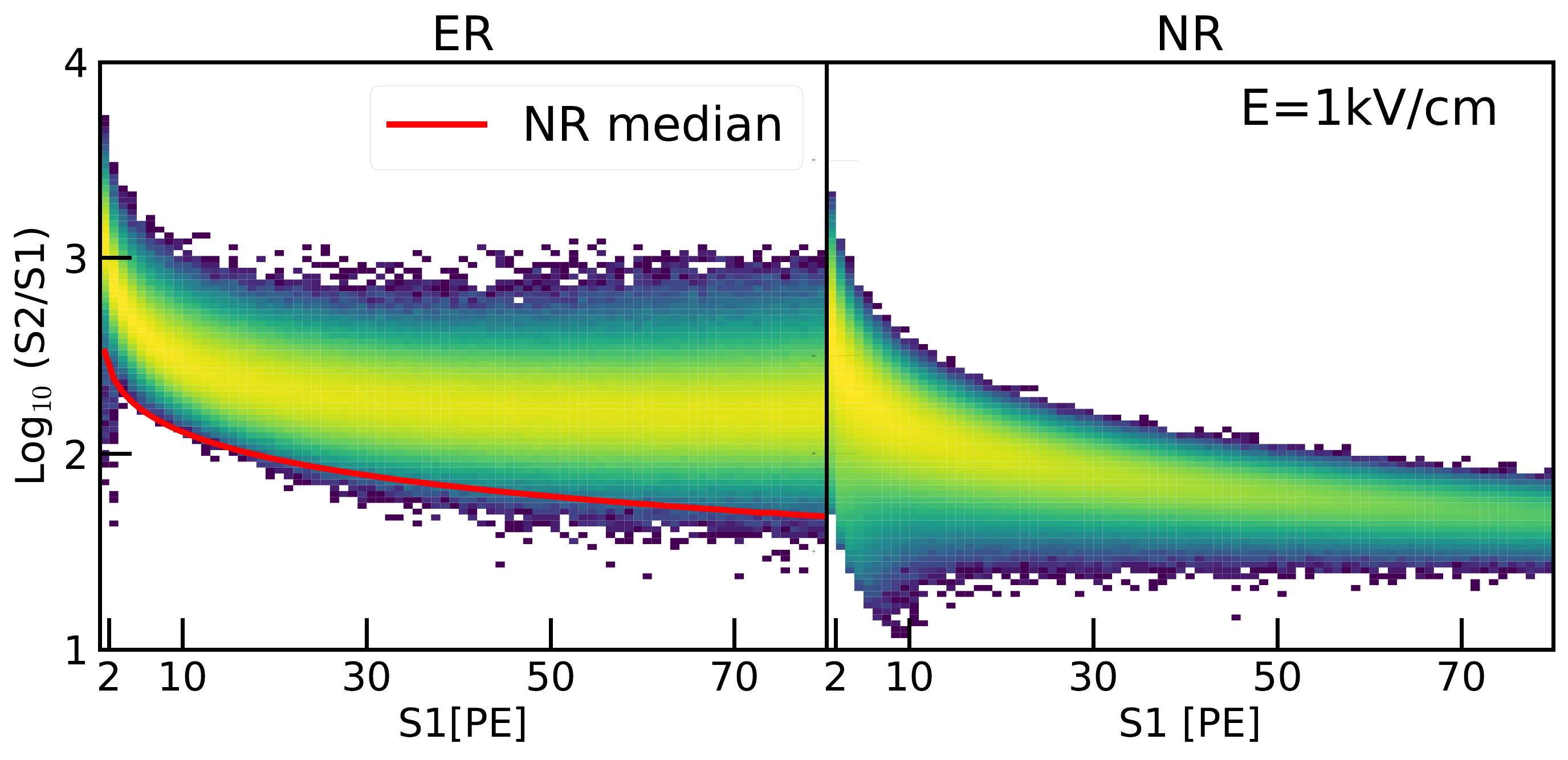}
    \caption{
    Distributions of simulated ER (left) and NR (right) events on Log$_{10}$(S2/S1) versus S1.
    The top, middle, and bottom panels show the simulated distributions under electric fields of 40, 200, and 1000\,V/cm, respectively.
    The red solid lines in figures in left column are the corresponding medians of NR distributions as a function of S1.
    }
    \label{fig:sim_er_nr_band}
\end{figure}

% simulation details and results
A toy simulation for ER-NR discrimination power in prototype GG-TPC is performed.
The simulation basically follows light and charge yield modelling in NESTv2~\cite{szydagis_m_2018_1314669}, which includes yields' dependences on electric field strength, incident energy, and recoil type (ER or NR).
The simulation also takes into account detector effects of photon detection, electron amplification, and double photo-electron (PE) emission effect of PMT photocathode~\cite{faham2015measurements} taken into account (for a reference of detector effect see~\cite{collaboration2019xenon1t}).
Other detector effects are minor and not included here.
The photon detection probability and electron amplification factor, so called $g_1$ and $g_2$, are assumed to be 0.1\,PE per photon and 20\,PE per drifted electron, respectively.
The standard deviation of g2 is assumed to be about 30\%.
Double PE probability is set to 20\% according to measurements~\cite{faham2015measurements}.
Fig.~\ref{fig:sim_er_nr_band} shows the distributions of simulated ER and NR events on Log$_{10}$(S2/S1) versus S1, under electric field of 40, 200, and 1000\,V/cm, respectively.
For each simulation shown in Fig.~\ref{fig:sim_er_nr_band}, 10\,M events are generated.
ER energy spectrum is assumed to be uniform from 0 to 70\,keV, and NR energy spectrum follows the one induced by a WIMP with mass of 200\,GeV.
In addition, an S2 threshold of 100\,PE is applied.

\begin{figure}[htp]
    \centering
    \includegraphics[width=0.48\textwidth]{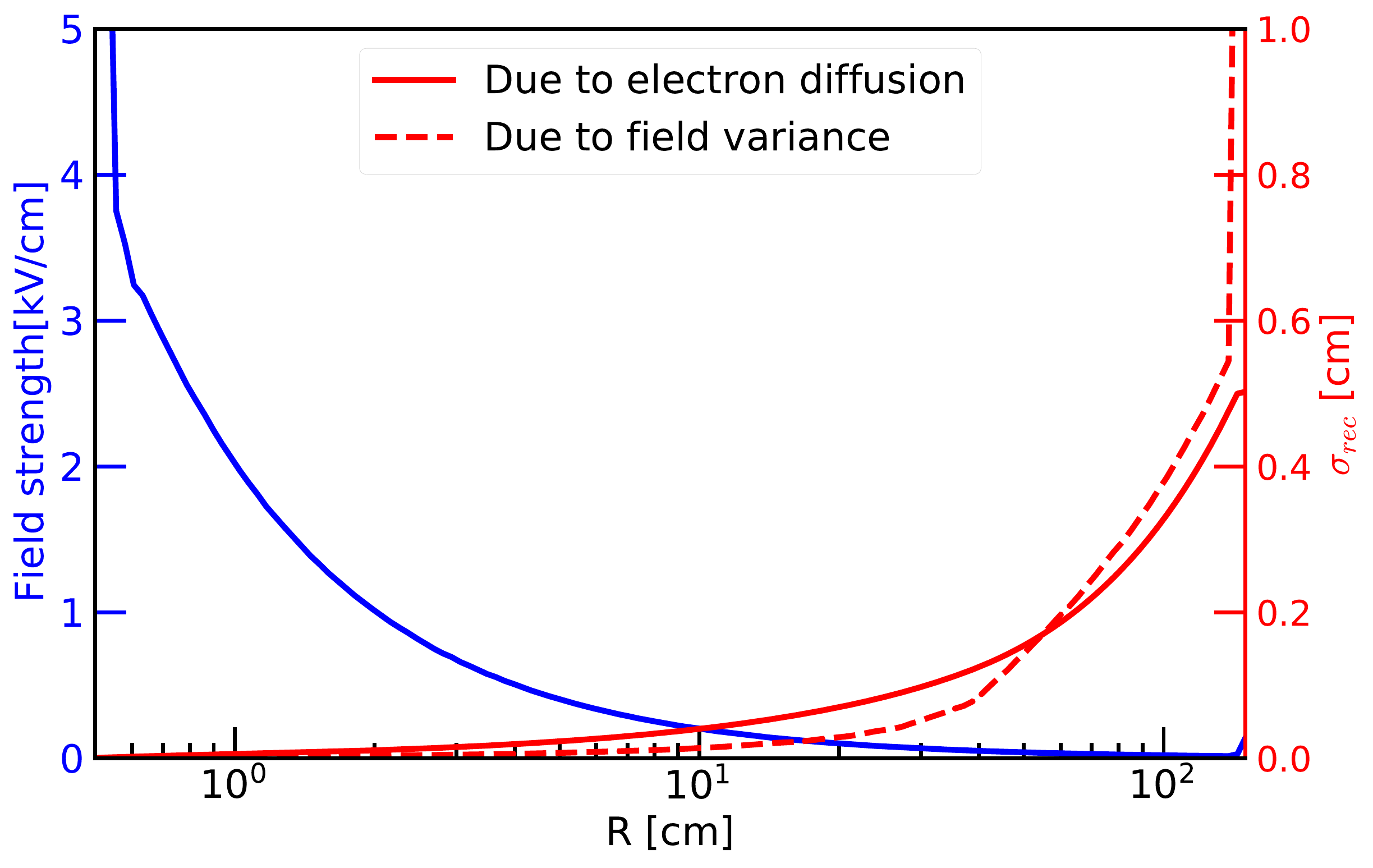}
    \includegraphics[width=0.48\textwidth]{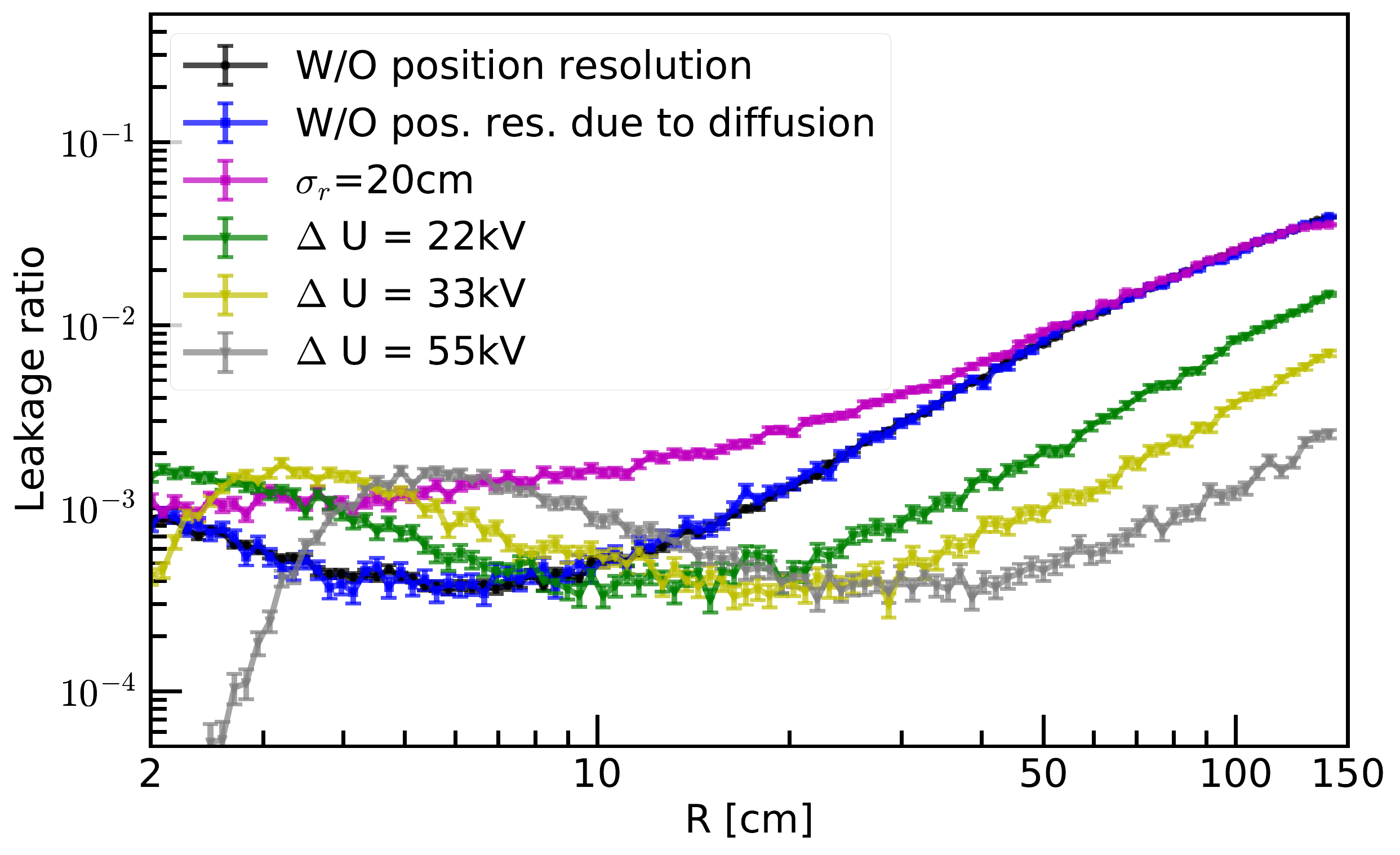}
    \caption{
    (Left) Blue solid line gives the electric field strength as a function of radial position R in prototype GG-TPC with gate and cathode voltages at 10\,kV and -1\,kV, respectively.
    Red solid and dashed lines give the R resolutions as a function of R, caused by longitude electron diffusion during drifting and field variance, respectively. 
    (Right) ER leakage ratio as a function of R.
    Black line gives the leakage ratio without considering R resolution, and blue line gives the one with realistic R resolution dominated by electron longitude diffusion.
    Magenta line gives the leakage ratio with a extreme R resolution of 20\,cm.
    Green, yellow and gray lines give leakage ratios with increased cathode-gate voltage difference (22, 33, and 55\,kV, respectively).
    The error bars represent the statistical uncertainties caused by finite simulation statistics.
    }
    \label{fig:diffusion_leakage_vs_r}
\end{figure}

% further talking about sim
NR distribution does not change much as electric field gets larger, but ER distribution changes considerably: both the S2/S1 ratio and its fluctuation gets larger at higher electric field.
In order to obtain a realistic estimation of leakage ratio (defined as the probability of ER events to appear under NR median on S2 versus S1 space) of GG-TPC, we consider the resolution of radial position reconstruction in GG-TPC is dominated by electron diffusion and field variance. 
Low radial position reconstruction resolution can cause the events at close radial positions (with different electric fields) are not distinguishable, reducing the ER-NR discrimination power.
We calculate the radial position resolution in prototype GG-TPC induced by field variance (results shown in Fig.~\ref{fig:sim_field_vs_R_largeTPC}) and electron diffusion, which is shown in left panel of Fig.~\ref{fig:diffusion_leakage_vs_r}.
In the calculation, we use the electron drift velocity measured in~\cite{yoshino1976effect} (linearly extrapolated below 50\,V/cm assuming a zero velocity at 0\,V/cm), and longitude electron diffusion coefficients as a function of field strength obtained by a global fit in~\cite{aalbers2018dark}.
To be conservative, field variances at different R positions are assumed to be fully correlated with each other.
The R resolution due to electron diffusion is dominant when radius of event is below about 50\,cm, and R resolution due to field variance gets bigger at larger radius (it can reach about 1.5\,cm near cathode).
%Left plot of Fig.~\ref{fig:diffusion_leakage_vs_r} shows the field strengths and radial position resolutions as a function of radial position for GG-TPC-P2.
Right plot of Fig.~\ref{fig:diffusion_leakage_vs_r} gives the leakage ratio as a function of radial position R, with the R resolution and field variance both taken into account in the calculation.

% discussion
In a GG-TPC with radius of 1.5\,meter and voltage difference between gate and cathode of 11\,kV, the ER leakage rate can range from about 3$\times$10$^{-4}$ to 0.04 depending on the event radial position.
The radial position reconstruction resolution does affect the ER rejection power, as shown by the magenta line in right plot of Fig.~\ref{fig:diffusion_leakage_vs_r} with a extreme low resolution of 20\,cm.
However GG-TPC's radial position resolution is dominated by electron diffusion, and, at maximum, reaches about 0.3\,cm.
It does not significantly reduce the ER rejection power in GG-TPC.
With the default field configuration (cathode at -1\,kV and gate at +10\,kV), the ER leakage ratio is worse than ones of past dual-phase LXe TPCs~\cite{xiao2015low, tan2016dark, aprile2012dark, baudis2018dark, akerib2017results}.
According to the toy simulation, further increasing the voltage difference between gate and cathode can improve the average ER rejection power considerably, as shown by green, yellow and gray lines in right plot of Fig.~\ref{fig:diffusion_leakage_vs_r} with voltage differences of 22, 33, and 55\,kV, respectively.
It is worth mentioning that NESTv2~\cite{szydagis_m_2018_1314669} that is used in toy simulation is derived using global data with electric field strength from about 80\,V/cm to 4\,kV/cm. 
The toy simulation results may have large uncertainty in the low and high field regions.
Besides, higher voltage difference between gate and cathode requires higher delivered voltage on them, bringing challenge to electrical feedthrough for detector system.
Therefore, the balancing between the ER rejection power and operational stable field configuration would need \textit{in-situ} adjusting, with dedicated calibrations using injected $^{220}$Rn source~\cite{aprile2017results, ma2020internal} and AmBe neutron source (or D-D generator)~\cite{aprile2019xenon1t, yan2021determination}, respectively, for ER and NR calibrations.

\section{Summary and Outlook}
\label{sec:summary}

% Summary 
We propose a new type of single phase TPC (GG-TPC), which has detector geometry similar to a Geiger counter and has the potential to be used as dark matter direct search detector in future experiment.
With a single wire as anode at central axis of cylindrical sensitive volume, the field strength near the anode wire can reach the threshold field strength for electroluminescence in LXe (420\,kV/cm) with moderate voltages applied.
Such GG-TPC conceptually can reduce the rate of isolated ionization signals which can affect the signal reconstruction and form accidental backgrounds.
GG-TPC has no liquid-gas surface so that it has no isolated ionization signals coming from delayed extraction of electrons~\cite{sorensen2017electron, sorensen2017two}.
Different from dual phase TPC which needs to minimize the voltage applied to anode since anode is in gas, GG-TPC can have positive high voltages applied to anode and gate, minimizing the cathode voltage.
Thus, GG-TPC can have minimal field strength around cathode wires, mitigating the electron emission from metal wire through Fowler-Nordheim effect~\cite{bodnia2021electric} due to the strong electric field between cathode and other detector material surface (such as PMT surface).
In addition, GG-TPC has less mechanic demands on electrode compared to a dual-phase TPC which can help to maintain the stable operation in large detector.
Table~\ref{tab:single_dual_comparison} shows a comparison of detector parameters between the GG-TPC discussed in this manuscript and a typical dual phase TPC with the same size.

\begin{table}[htp]
    \centering
    \begin{tabular}{c||c|c}
         & GG-TPC & dual phase TPC \\
    \hline\hline
    Anode voltage [kV]     & +40 & +5 \\
    Gate voltage [kV]     & +10 & 0 \\
    Cathode voltage [kV]     & -1 & -11 \\
    Avg. ER leakage ratio      & 0.023  & 0.009 \\
    Wire coverage [cm$^2$]     & 2700 & 5655 \\
    Drift length [cm]     & 150 & 300 \\
    \hline\hline
    \end{tabular}
    \caption{Comparison of some parameters between GG-TPC and traditional dual phase TPC with same voltage difference between gate and cathode, wire size (200\,$\mu m$), and detector dimension (3\,meter in height and diameter).
    Electrodes of dual phase TPC is assumed to be wire plane with pitch of 5\,mm.
    }
    \label{tab:single_dual_comparison}
\end{table}

% feasibility check results
On the other hand, the challenges are that GG-TPC has non-uniform field in sensitive volume along radial position, proportional lights for a single electron has time profile close to prompt scintillation lights, and position reconstruction.
Field simulation shows that in most part of GG-TPC's active volume, the standard deviation of field strength can reach approximately below 1\% level.
It is able to reach good signal quality with correction for GG-TPC's radial dependence of field.
The time duration of SE S2s in GG-TPC is similar to that of prompt S1, but do have different pulse shape.
We have performed simple MLP-based classification between S1s and SE S2s in GG-TPC using toy MC data.
Results show that PSD classification power can reach about 90\% when the light collection of S2 is sufficiently large to have SE S2 at $>$40 hits.
The digitizer sampling rate plays a minor role in the classification performance.
Radial positions of events in GG-TPC are reconstructed using drift time, thus having high resolution.
Angular positions are with low resolution due to Rayleigh scattering and finite number of gate wires in large GG-TPC. 
However angular information of events is not critical in an axial symmetrical detector.
$Z$ positions of events in GG-TPC are of great importance in terms of rejecting external gamma-rays and surface backgrounds.
Using hit pattern on detector side, the $Z$ reconstruction quality is not good ($\sim$20\,cm near top/bottom) due to the fact that PMTs are far away from detector central axis where S2s are produced.
Instrumenting light sensors on top and bottom surfaces can greatly improve the resolution of $Z$ position reconstruction in GG-TPC, both through the top-bottom asymmetry and reconstruction using hit pattern.
In regions near top/bottom planes, the best $Z$ resolution estimated can reach sub-cm level for a hit number of $\sim$1000 hits.
The ER rejection power of GG-TPC is not significantly affected by the radial position resolution.
We perform a toy signal response simulation and show that ER leakage ratio can basically reach about 3$\times$10$^{-4}$ to 0.04 in GG-TPC with 1.5\,meter and voltage difference of 11\,kV between gate and cathode.
Further \textit{in-situ} improvement of ER rejection power can be done by increasing voltage of gate or cathode.

% Outlook (mainly about future challenges)
In summary, GG-TPC is conceptually a promising technique that can be used in future dark matter direct search experiment with low isolated S2 signals and low accidental pileup backgrounds.
However, realistic mechanical precision may affect GG-TPC's performance which needs R\&D work to demonstrate.
For example, the surface smoothness of anode wire may violate the axial symmetry of GG-TPC, which needs extra treatment.
Also, it is challenge to place light sensors at top/bottom planes in GG-TPC.
If PMTs are used as light sensors, shielding electrodes are needed to protect PMTs and to avoid field distortion in top/bottom regions.
Besides this, special light collection techniques, like light guides coupled with PMTs, can also be explored.
In addition, silicon PMTs (SiPM) can be used as light sensors at top/bottom of GG-TPC instead of traditional PMTs, to mitigate the field distortion problem because SiPM's surface is not at high voltage.
However SiPM needs dedicated R\&D to demonstrate its own feasibility in dark matter direct search experiments, in terms of quantum efficiency, dark rate and saturation.
Depending on the size, dark rate and saturation charge of SiPM, it is also option to use both SiPMs and PMTs as light sensors for GG-TPC, such as using SiPMs on top/bottom for position reconstruction and PMTs at cylinder side for light collection. 

% Other light sensors, such as SiPMT, and special light collection techniques, such as light guides coupled with PMTs, need to be explored.

% \acknowledgments

% % We thank Kaixuan Ni and Yuehuan Wei for helpful discussions. 
% Q.L. is supported by One Thousand Talent Program for Young Scholars.

%\bibliographystyle{apsrev}
%\bibliography{reference.bib}

\end{document}